\documentclass[journal]{IEEEtran}

\usepackage{amssymb}
\usepackage{amsmath}
\usepackage{bm} 
\usepackage{relsize} 
\usepackage{graphicx}
\usepackage{subfigure}
\usepackage{multicol}
\usepackage{url}
\usepackage{cite}
\usepackage{enumerate}
\usepackage{epstopdf}
\usepackage[ruled, procnumbered]{algorithm2e}
\usepackage{color}
\usepackage{hyperref}
\usepackage{cleveref}

\begin{document}

\title{\huge{Flexible Energy Management Protocol for Cooperative EV-to-EV Charging}}

\author{Rongqing~Zhang,~\IEEEmembership{Member,~IEEE,} Xiang~Cheng,~\IEEEmembership{Senior~Member,~IEEE} and~Liuqing~Yang,~\IEEEmembership{Fellow,~IEEE,}
\thanks{Rongqing Zhang is with the State Key Laboratory of Advanced Optical Communication Systems and Networks, School of Electronics Engineering and Computer Science, Peking University, Beijing, China, and also with the Department of Electrical
\& Computer Engineering, Colorado State University, CO, USA (Email: rongqing.zhang@colostate.edu).}
\thanks{Xiang Cheng is with the State Key Laboratory of Advanced Optical Communication Systems and Networks, School of Electronics Engineering and Computer Science, Peking University, Beijing, China, and also with the Shenzhen Research Institute of Big Data, Shenzhen, Guangdong, China (Email: xiangcheng@pku.edu.cn).}
\thanks{Liuqing Yang is with the Department of Electrical \& Computer Engineering, Colorado State University, CO, USA, and also with the Shenzhen Research Institute of Big Data, Shenzhen, Guangdong, China (Email: lqyang@engr.colostate.edu).}
}


\renewcommand{\thepage}{}
{} \maketitle
\pagenumbering{arabic}\setcounter{page}{1}

\begin{abstract}

In this paper, we investigate flexible power transfer among electric vehicles (EVs) from a cooperative perspective in an EV system. First, the concept of cooperative EV-to-EV (V2V) charging is introduced, which enables active cooperation via charging/discharging operations between EVs as energy consumers and EVs as energy providers. Then, based on the cooperative V2V charging concept, a flexible energy management protocol with different V2V matching algorithms is proposed, which can help the EVs achieve more flexible and smarter charging/discharging behaviors. In the proposed energy management protocol, we define the utilities of the EVs based on the cost and profit through cooperative V2V charging and employ the bipartite graph to model the charging/discharging cooperation between EVs as energy consumers and EVs as energy providers. Based on the constructed bipartite graph, a max-weight V2V matching algorithm is proposed in order to optimize the network social welfare. Moreover, taking individual rationality into consideration, we further introduce the stable matching concepts and propose two stable V2V matching algorithms, which can yield the EV-consumer-optimal and EV-provider-optimal stable V2V matchings, respectively. Simulation results verify the efficiency of our proposed cooperative V2V charging based energy management protocol in improving the EV utilities and the network social welfare as well as reducing the energy consumption of the EVs.

\end{abstract}

\begin{keywords}
Electric vehicles, cooperative V2V charging, matching theory, energy management.
\end{keywords}

\IEEEpeerreviewmaketitle


\section{Introduction}%

With ever increasing concerns on environmental issues and clean energy, electric vehicles (EVs) have attracted more and more attention from governments, industries, and costumers \cite{R1-1}. EVs are regarded as one of the most effective strategies to reduce the oil dependence and gas emission, and to increase the efficiency of energy conversion \cite{EV-Chan-2007,EV-Issues-2011,XCheng-IS-2016}. When integrated with the power grid based on charging and/or discharging operations, EVs become energy storage units, and can not only serve as a transportation tool but also act as controllable loads and distributed sources for the power grid \cite{R1-2,RQ-EV-ComMag-2016}.

On the one hand, the fast development of EVs brings a significant new load on the current power system \cite{Lopes-2011}. Without efficient control strategies, the EV charging process may overload the power grid at peak hours, especially in residential communities. On the other hand, EVs can benefit the power grid as a flexible load through smart charging/discharging scheduling to reduce the peak load and shape the load profile. Therefore, in the literature, with the concept of demand side management (DSM) \cite{Palensky-2011}, many works \cite{Chow-2014,NS-2012,TYN-2014,YDZLX-2014,KWZA-2016,BF-2014} have focused on the charging/discharging scheduling and energy management protocols to control and optimize the charging process for EVs integrated with the power grid.

In \cite{NS-2012}, the authors investigated a smart charging and discharging process for multiple EVs parked in a building garage to optimize the energy consumption profile of the building. An energy charging and discharging scheduling game was formulated to control the charging and discharging behaviors of the EVs in order to minimize the peak load and the total energy cost. In \cite{YDZLX-2014}, the authors investigated the charging and discharging cooperation between the power grid and the EVs, and formulated and resolved this cooperation in the framework of a coalitional game to make the EVs have the incentive to charge in load valley and discharge in load peak. In \cite{Chow-2014}, the authors developed a cooperative distributed algorithm for charging control of EVs through peer-to-peer coordination of charging stations in a distributed fashion. In \cite{TYN-2014}, a price scheme considering both base price and demand fluctuation in the demand response was proposed and then a distributed optimization algorithm based on the alternating direction method of multipliers was developed to optimize the demand side management problem for the future smart grid with EVs and renewable distributed generators. In \cite{KWZA-2016}, the authors proposed a centralized charging strategy of EVs under the battery swapping scenario by considering optimal charging priority and charging location based on spot electric price, in order to minimize total charging cost, as well as to reduce power loss and voltage deviation of the power systems. In \cite{BF-2014}, motivated by the demand for an efficient EV charging model in the bidirectional vehicle-to-grid (V2G) integration, a convex quadratic programming framework was developed for the charge pattern optimization of EVs under time-varying electricity price signals.
Although various optimization methods as well as game theory models have been employed to design different EV charging and energy management protocols in existing work, the current researches are still limited to the interactions and power transfer between EVs and the power grid.

In recent years, Internet of Energy (IoE), as an important part of Internet of Things, is regarded as a promising concept for future energy system. In IoE, renewable-energy power plants, transmission links, electrical meters, appliances, and the moving EVs will be able to talk to each other in real time about the electrical loads and energy prices and share power with each other if demanded \cite{Huang-2011}. Just like the information Internet forever changed the way information is made, shared, and stored, IoE will also change the way we produce, distribute, and store energy. IoE is envisioned as a smart architecture that enables flexible energy sharing among the involved units. The mobile and energy storage features make the EVs play an important role in increasing the flexibility and possibilities of power transfer in the IoE.


Most recently, some works \cite{Book-EV-2016,Gerding-2016} have proposed to investigate vehicle-to-vehicle charging strategies, which can offer more flexible charging plans for gridable EVs in order to offload the EV charging loads from the electric power systems. However, designing an effective and efficient online vehicle-to-vehicle charging strategy remains an open issue. In this paper, we investigate the flexible power transfer among EVs from a cooperative perspective in the energy Internet based EV system. First, we introduce the concept of cooperative EV-to-EV (V2V) charging, which enables active cooperation through charging/discharging operations between EVs as energy consumers and EVs as energy providers. Then, based on the cooperative V2V charging concept, we propose a flexible energy management protocol with different V2V matching algorithms, which can help the EVs achieve more flexible and smarter charging/discharging behaviors. Simulation results verify the efficiency of the proposed cooperative V2V charging based energy management protocol in improving the EV utilities as well as the network social welfare. Moreover, the simulation results also indicate that the energy consumption of the EVs can be reduced effectively with the proposed cooperative V2V charging based energy management protocol.

The main contributions of this paper can be summarized as follows.
\begin{enumerate}[1.]
  \item We first provide a developed concept based on the V2V operation of V2X concept introduced in \cite{V2X-Liu-2013}, termed as cooperative V2V charging, which describes the power flow connection among different EVs in a cooperative charging/discharging manner. Cooperative V2V charing can enable direct EV-to-EV power transfer through active cooperation between EVs as energy consumers and EVs as energy providers. Through the cooperative V2V charging, the charging/discharging behaviors of EVs can be performed in a more flexible and smarter manner.

  \item Based on cooperative V2V charging, we propose a novel and flexible energy management protocol. In the proposed energy management protocol, EVs as energy consumers and EVs as energy providers can send their real-time individual information and energy trading requests to the data control center via mobile apps or on-board apps under the support of mobile Internet and intelligent transportation system (ITS). According to the collected information, the data control center will make smart charging/discharging decisions for the involved EVs through effective and efficient V2V matching, which are beneficial to both EVs as energy consumers and EVs as energy providers.

  \item In order to achieve the optimal V2V matching in terms of network social welfare, we construct a weighted bipartite graph, in which EVs as energy consumers and EVs as energy providers are formulated as two partite sets. Based on the constructed bipartite graph, we propose a max-weight V2V matching algorithm, which can obtain the maximum weight V2V matching with low complexity.

  \item Although the proposed max-weight V2V matching algorithm can output an optimized V2V matching in terms of network social welfare, the obtained V2V matching is not a stable one, which means the matched EVs may have an incentive to deviate from the matching. Therefore, taking the individual rationality of each involved EV into consideration, we further introduce the stable matching concepts and propose two stable V2V matching algorithms, i.e., the EV-consumer-oriented stable V2V matching algorithm and the EV-provider-oriented stable V2V matching algorithm, which yield the EV-consumer-optimal and EV-provider-optimal stable V2V matchings, respectively.
\end{enumerate}

The remainder of this paper is organized as follows. In Section II, the system model is described and the cooperative V2V charging concept is introduced. In Section III, based on the cooperative V2V charging concept, we provide a flexible energy management protocol to achieve effective and smart charging/discharging cooperation for the EVs. In Section IV, by employing the matching theory, we construct a bipartite graph to indicate the relationships between EVs as energy consumers and EVs as energy providers and further propose three efficient V2V matching algorithms. Simulations are provided in Section V and the conclusions are given in Section VI.

\section{System Model}%

\subsection{System Description}

\begin{figure}[!t]
\centering
\includegraphics[width=3.6in]{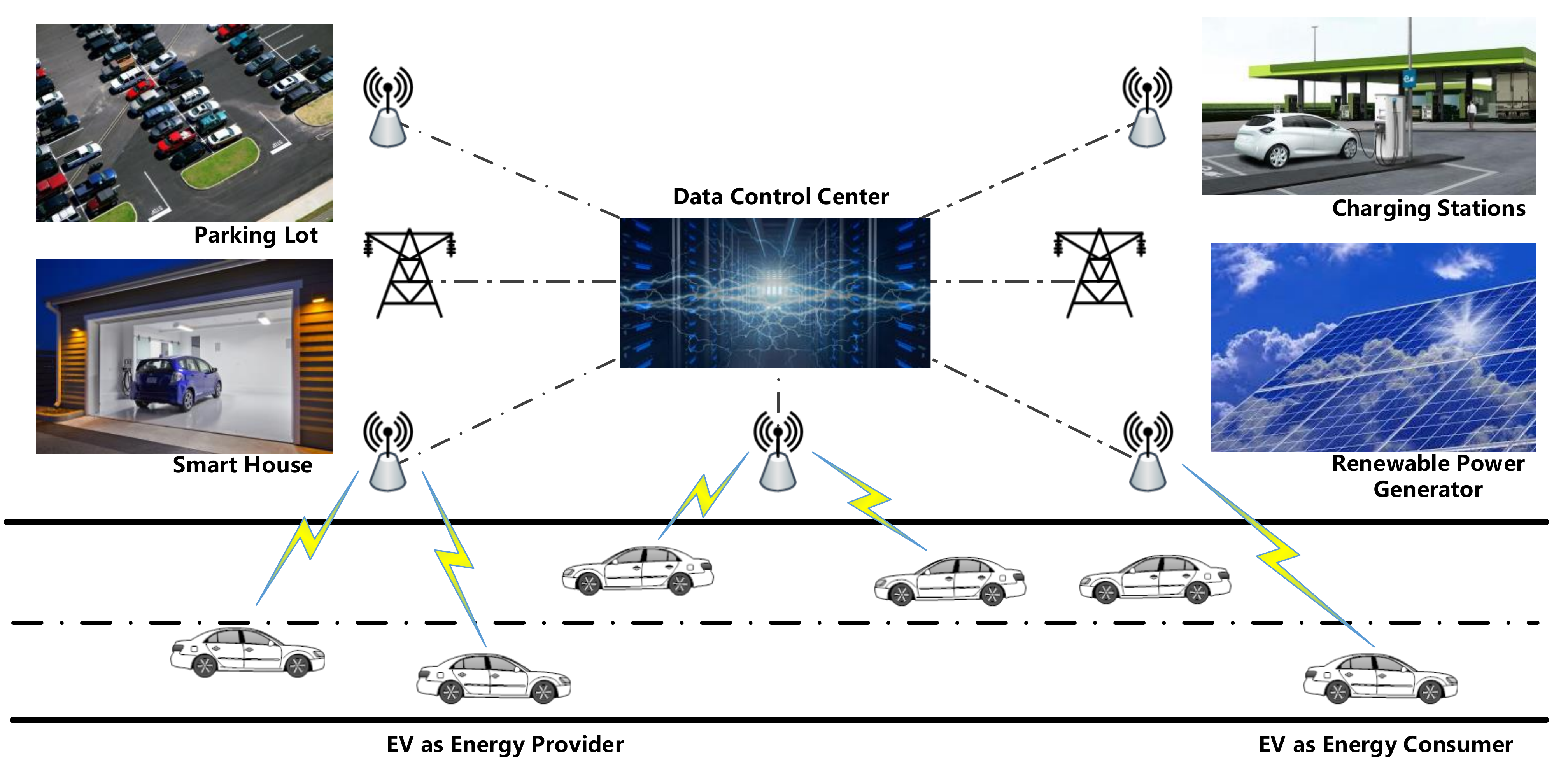}
\caption{An illustration of our investigated IoE based EV system.} \label{Fig-System_Model}
\end{figure}

As illustrated in Fig.~\ref{Fig-System_Model}, we consider an IoE based EV system, mainly comprising the EVs, the smart houses, the charging stations, the power/communication infrastructures, and a data control center. Each EV is equipped with an bidirectional charger and thus can perform both charging and discharging behaviors. The moving EVs in the investigated system can be divided into three categories: 1) EVs that demand power act as energy consumers, denoted by $\mbox{EV}_i^C$, $i=1,2,\ldots,N$; 2) EVs that have extra power act as energy providers, denoted by $\mbox{EV}_j^P$, $j=1,2,\ldots,K$; 3) EVs that are not interested to participate in any current energy trading. For presentation convenience, we denote $\mathcal{N}\triangleq\left\{1,2,\ldots,N\right\}$ and $\mathcal{K}\triangleq\left\{1,2,\ldots,K\right\}$. The data control center is connected to all the distributed power and information infrastructures and can collect the real-time information about the nearby charging stations, smart houses, parking lots, and the EVs via 5G-enabled IoT \cite{XCheng-IS-2017}.

\subsection{Cooperative V2V Charging}

In this section, we propose a developed concept based on the V2V operation of V2X concept \cite{V2X-Liu-2013}, termed as cooperative V2V charging, which describes the power flow connection among different EVs in a cooperative charging/discharging manner. Cooperative V2V charging can enable direct EV-to-EV power transfer through active cooperation among EVs at the energy level. Based on cooperative V2V charging, the charging/discharging behaviors of EVs can be performed in a more flexible and smarter manner.

Cooperative V2V charging is beneficial to both EVs as energy consumers and EVs as energy providers, leading to a win-win energy trading situation. As for EVs as energy consumers, currently, most EVs get charged at the charging stations or after going back home. If a moving EV demands power before it can arrive at the destination, it will have to drive to a nearby charging station to get charged first. However, the current deployment of charging stations is still far from sufficient. Moreover, the nearby charging station may be in a different direction deviating from the EV's original route to the destination. This causes inconvenience and extra energy consumption for EVs as energy consumers. As an additional feasible charging option, cooperative V2V charging can make the charging behaviors of EVs as energy consumers more flexible and smarter, and thus reduce the drivers' anxiety since their EVs can get charged more easily. As for EVs as energy providers, the EV drivers can make profits through cooperative V2V charging based energy trading with their spare time and surplus power. Especially, with the increasing penetration of renewable energy resources (RESs) such as solar panels in residential houses, many households would have surplus power generated by RESs at a low cost, leading to considerable incentives and motivations on individual energy trading. Even taking the potential cost (e.g., the battery lifetime loss) into consideration, there is still a profit margin for EVs as energy providers to achieve cooperative V2V charging based energy trading with their stored low-cost surplus power. Besides, cooperative V2V charging can also offload the heavy load of the power grid due to the dramatically increasing penetration of EVs in daily life.

Currently, a feasible way to realize V2V power transfer among different EVs is through the V2V framework described in \cite{V2X-Liu-2013}, where an aggregator is employed for coordinated control of grouping EVs for charging and discharging. The aggregator behaves as a control device that collects all the information about the EVs and the grid status and then executes the V2V power transfer. Since these aggregators do not need to pull in power from the power grid to operate the V2V power transfer, they would be much cheaper and more easily deployed than the charging stations. For instance, such aggregators can be widely deployed in various communities or public parking lots. One can also envision that in the future IoE, the power transfer among EVs may be achieved via a single charging cable connecting EVs directly or even in a wireless and mobile manner (i.e., wireless V2V power transfer). This will make the charging/discharging among EVs more easily and conveniently. Then, with cooperative V2V charging, EVs will be able to get charged anytime anywhere in the future.

\section{Cooperative V2V Charging Based Energy Management Protocol}%

In this section, based on the cooperative V2V charging concept, we propose a flexible energy management protocol for the EVs in the investigated system. In the designed protocol, EVs that demand power can send the charging requests to the data control center via mobile apps or on-board apps under the support of 5G-enabled Internet of Vehicles (IoV) and intelligent transportation system (ITS) \cite{XCheng-2014,Fanhui-CL-2017,XCheng-D2D-2015,RQ-TVT-2013}. Meanwhile, EVs that have extra power can also send their real-time individual information to the data control center and wait for the energy trading decisions. Then, based on the collected information, the data control center will make smart charging/discharging decisions for the involved EVs through effective and efficient V2V matching, which are beneficial to both sides (i.e., EVs as energy consumers and EVs as energy providers). In the following, we will detail the proposed cooperative V2V charing based energy management protocol design.

\subsection{EV Utility Definition}

First, we need to define the utilities of the EVs as energy consumers and energy providers based on their cost/profits through potential energy trading and the corresponding energy cost for driving to the selected trading spot.

\subsubsection{EV as an Energy Consumer}

The utility of $\mbox{EV}_i^{C}$, $i\in\mathcal{N}$ as an energy consumer is defined as
\begin{align} \label{Utility_EV_C}
U_i^C \left(\mbox{EV}_j^{P}\right)  = - p_t a_i^C  -  \mbox{Cost} \left(\mbox{EV}_i^{C},\mbox{EV}_j^{P}\right)
\end{align}
where $p_t$ is the unit power trading price, $a_i^C$ represents the requested power amount, and $\mbox{EV}_j^P$ is the potential paired energy provider for $\mbox{EV}_i^{C}$. In general, the electricity buying price $p_b$ set by the power grid for the EVs to trade their surplus power is often considerably lower than the electricity selling price $p_s$ for the EVs to get charged \cite{Three-Party-SG-2015}. Based on this, the unit power trading price $p_t$ can be set between the electricity buying and selling prices of the power grid. Therefore, EVs as energy providers can sell their surplus power at a higher price, and EVs as energy consumers can also buy their requested power at a lower price, compared with energy trading through discharging and charging directly with the power grid. In practical applications, the unit power trading price $p_t$ may also vary based on the current information collected at the data control center.


As a preference baseline, we also define the utility of $\mbox{EV}_i^{C}$ when getting charged at the charging stations as
\begin{align} \label{Utility_EV_C_CS}
U_i^C\left(\mbox{CS}\right)  = - p_s a_i^C  - \mbox{Cost} \left(\mbox{EV}_i^{C},\mbox{CS}\right)
\end{align}
where $\mbox{CS}$ denotes the nearest charging station for $\mbox{EV}_i^{C}$ and $p_s$ is the electricity selling price set by the power grid, that is, the power trading price between the charging station and the EVs as energy consumers.

Note that $\mbox{Cost} \left(\mbox{EV}_i^{C},\mbox{EV}_j^{P}\right)$ and $\mbox{Cost} \left(\mbox{EV}_i^{C},\mbox{CS}\right)$ denote the energy cost for $\mbox{EV}_i^{C}$ to drive to the selected parking lot to achieve power transfer with $\mbox{EV}_j^{P}$ and to drive to the nearest charging station to get charged, respectively, which can be given as
\begin{align} \label{Cost_EVc_PL}
\mbox{Cost} \left(\mbox{EV}_i^{C},\mbox{EV}_j^{P}\right) = p_t \times \beta_{i}^{C} \times \mbox{Dis}\left(\mbox{EV}_i^{C},\mbox{PL}\right)
\end{align}
and
\begin{align} \label{Cost_EVc_CS}
\mbox{Cost} \left(\mbox{EV}_i^{C},\mbox{CS}\right) = p_s \times \beta_{i}^{C}  \times \mbox{Dis}\left(\mbox{EV}_i^{C},\mbox{CS}\right)
\end{align}
where $\beta_{i}^{C}$ is the moving energy cost per km for $\mbox{EV}_i^{C}$, $ \mbox{Dis}\left(x,y\right)$ represents the driving distance between $x$ and $y$, and $\mbox{PL}$ denotes the selected parking lot for $\mbox{EV}_i^{C}$ to achieve power transfer with $\mbox{EV}_j^{P}$. Note that here the energy cost for $\mbox{EV}_i^{C}$ to get charged at the charging station is valued by the electricity selling price $p_s$ of the power grid.

\subsubsection{EV as an Energy Provider}

The utility of $\mbox{EV}_j^{P}$, $j\in\mathcal{K}$ as an energy provider is defined as
\begin{align} \label{Utility_EV_P}
U_j^P \left(\mbox{EV}_i^{C}\right) &= p_t a_i^C - p_0 a_i^C/\eta - \mbox{Cost}\left(\mbox{EV}_j^{P},\mbox{EV}_i^{C}\right) \nonumber \\
&- \mbox{Time}\left(\mbox{EV}_j^P,\mbox{EV}_i^C\right) - \Phi \left(\mbox{EV}_j^P,\mbox{EV}_i^C\right)
\end{align}
where $p_t$ and $p_0$ are the current trading price and the original cost price per unit power, respectively, $\eta$ represents the V2V power transfer efficiency, $\Phi \left(\mbox{EV}_j^P,\mbox{EV}_i^C\right)$ represents the amortized cost to value the battery degradation per each V2V power transfer. $\mbox{Cost} \left(\mbox{EV}_j^{P},\mbox{EV}_i^{C}\right)$ and $\mbox{Time}\left(\mbox{EV}_j^P,\mbox{EV}_i^C\right)$ denote the energy cost and the time cost for $\mbox{EV}_j^{P}$ to drive to the selected parking lot to achieve power transfer with $\mbox{EV}_i^{C}$, respectively, which can be given as
\begin{align} \label{Cost_EVp_PL}
\mbox{Cost} \left(\mbox{EV}_j^{P},\mbox{EV}_i^{C}\right) = p_t \times \beta_{j}^{P} \times \mbox{Dis}\left(\mbox{EV}_j^{P},\mbox{PL}\right)
\end{align}
and
\begin{align}
\mbox{Time}\left(\mbox{EV}_j^P,\mbox{EV}_i^C\right) = \theta_j^P \left(\frac{\mbox{Dis}\left(\mbox{EV}_j^P,\mbox{PL}\right)}{v_j^P} + \tau a_i^C/\eta \right)
\end{align}
where $\beta_j^P$ is the energy cost per km for $\mbox{EV}_j^{P}$, $\theta_j^P$ represents the value of time for $\mbox{EV}_j^P$, $v_j^P$ is the velocity of $\mbox{EV}_j^P$, and $\tau$ denotes the V2V power transfer speed per unit of power. Here we assume that the current surplus power amount of $\mbox{EV}_j^{P}$ for energy trading denoted by $a_j^P$ satisfies $a_j^P\geq a_i^C$.

As for the time cost, we would like to point out that how to value such a time cost objectively is difficult in practical applications, since the value of time for different people would be quite different and highly subjective. Similar to Uber drivers, the EV drivers who send their information to the control center to act as energy providers may mostly be the ones with some spare time and willing to make profits with their surplus power via energy trading. Therefore, in our opinion, the effect of the time cost should be much smaller than that of the energy cost. As for the amortized cost due to battery degradation, the value of $\Phi$ in practical applications can be calculated based on the information reported by the EVs and stored in the data control center. Referring to \cite{Battery-Degradation}, the battery degradation cost model takes both the battery replacement cost and the battery wear-out via energy processing into consideration, and then $\Phi \left(\mbox{EV}_j^P,\mbox{EV}_i^C\right)$ can be given as
\begin{align}
\Phi \left(\mbox{EV}_j^P,\mbox{EV}_i^C\right) = \phi_j \times D_j \times a_{i}^C
\end{align}
where $\phi_j$ represents the battery replacement cost of $\mbox{EV}_j^P$, and $D_j$ is the capacity degradation coefficient.

Since the electricity buying price $p_b$ set by the power grid for EVs as energy providers to trade their surplus power is lower than the power trading price $p_t$ via cooperative V2V charging, EVs as energy providers would prefer to trade with EVs as energy consumers instead of the power grid, if a positive utility can be achieved.

\begin{figure*}[!t]
\centering
\includegraphics[width=5.0in]{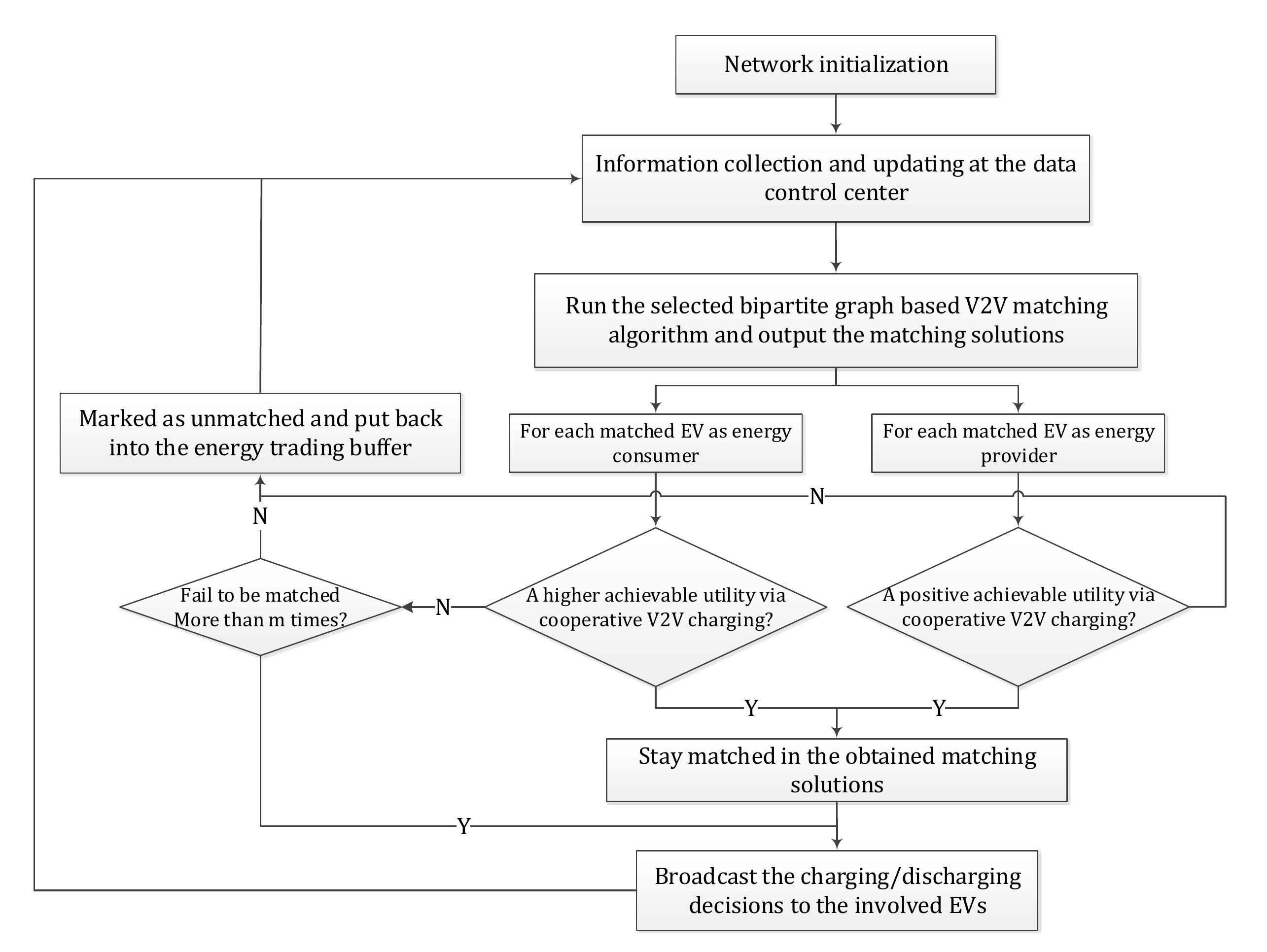}
\vspace{-3mm}
\caption{Flow chart of the protocol procedure at the data control center.} \label{Fig-Protocol_Procedure}
\vspace{-3mm}
\end{figure*}

\subsection{Flexible Energy Management Protocol Design}


Based on the cooperative V2V charging concept, we provide a flexible energy management protocol for the EVs in this section. The flow chart of the procedure of the proposed energy management protocol is given in Fig.~\ref{Fig-Protocol_Procedure}. The data control center acts as the central controller for the energy management process. During the procedure, the data control center collects and updates the real-time information via mobile Internet periodically. The collected information at the data control center includes the real-time location and moving information from EVs, the location information of the nearby charging stations, smart houses, and parking lots, the charging request and demanded power amount from EVs as energy consumers, the available trading power amount from EVs as energy providers, and the real-time unit electricity price from the charging stations. Based on the collected information, the data control center performs a selected bipartite graph based V2V matching algorithm to obtain an efficient and effective V2V matching and help the EVs make smart charging/discharging decisions.

Different from typical game theory based methods, in which players always require information on other players' actions to take their own best responses during the converging process, matching theory based algorithms can avoid frequent information exchange among EVs and thus reduce the system communication overhead. Besides, most game-theoretic solutions only investigate one-sided or unilateral stability notions \cite{MatchingTheory-Han-2014}, whereas stable matching algorithms can achieve two-sided stability for both the EVs as energy consumers and EVs as energy providers based on their respective preferences, which are more practical and suitable for the formulated assignment problem between two distinct sets of players. Therefore, in this paper, we employ matching theory to design the V2V matching algorithms. In Section IV, we will propose three bipartite graph based V2V matching algorithms, that is, the max-weight V2V matching algorithm that can achieve optimized network social welfare, the EV-consumer-oriented stable V2V matching algorithm that can achieve EV-consumer-optimal stable energy trading, and the EV-provider-oriented stable V2V matching algorithm that can achieve EV-provider-optimal stable energy trading. The selection of these three V2V matching algorithms in the proposed energy management protocol depends on the current criterion in the EV energy trading market.


Note that during the V2V matching process, the data control center will automatically choose a best available parking lot for each potential paired EV based on the stored parking lots information. After the V2V matching process, for each matched EV as energy consumer, the achievable utility through the cooperative V2V charging will be checked whether to be larger than the utility when getting charged at a nearby charging station. If not, the corresponding matched EV pair will be marked as unmatched and put into the energy trading buffer again. Similarly, for each matched EV as energy provider, if the achievable utility through the cooperative V2V charging is not a positive value, the corresponding matched EV pair will also be marked as unmatched and put into the energy trading buffer again. If an EV as an energy consumer fails to be matched for more than $m$ times, the data control center will feedback a failure-matched notice, which means it is a better option for the EV to get charged at the nearby charging stations. If a cooperative V2V charging deal is finally reached, the two involved EVs will perform the power transfer at a nearby parking lot selected by the data control center.

\section{Efficient V2V Matching Algorithms}%

In the proposed cooperative V2V charging based energy management protocol, V2V matching is the core process, which determines the efficiency of energy trading between EVs as energy consumers and EVs as energy providers. In order to achieve efficient and effective V2V matching, in this section, we first construct a bipartite graph to indicate the relationships between EVs as energy consumers and EVs as energy providers, and then based on the constructed bipartite graph, we further propose three V2V matching algorithms by employing matching theory \cite{MatchingTheory-Han-2014}, i.e., the max-weight V2V matching algorithm, the EV-consumer-oriented stable V2V matching algorithm, and the EV-provider-oriented stable V2V matching algorithm.

\subsection{Bipartite Graph Construction}

\begin{figure}[!t]
\centering
\includegraphics[width=3.6in]{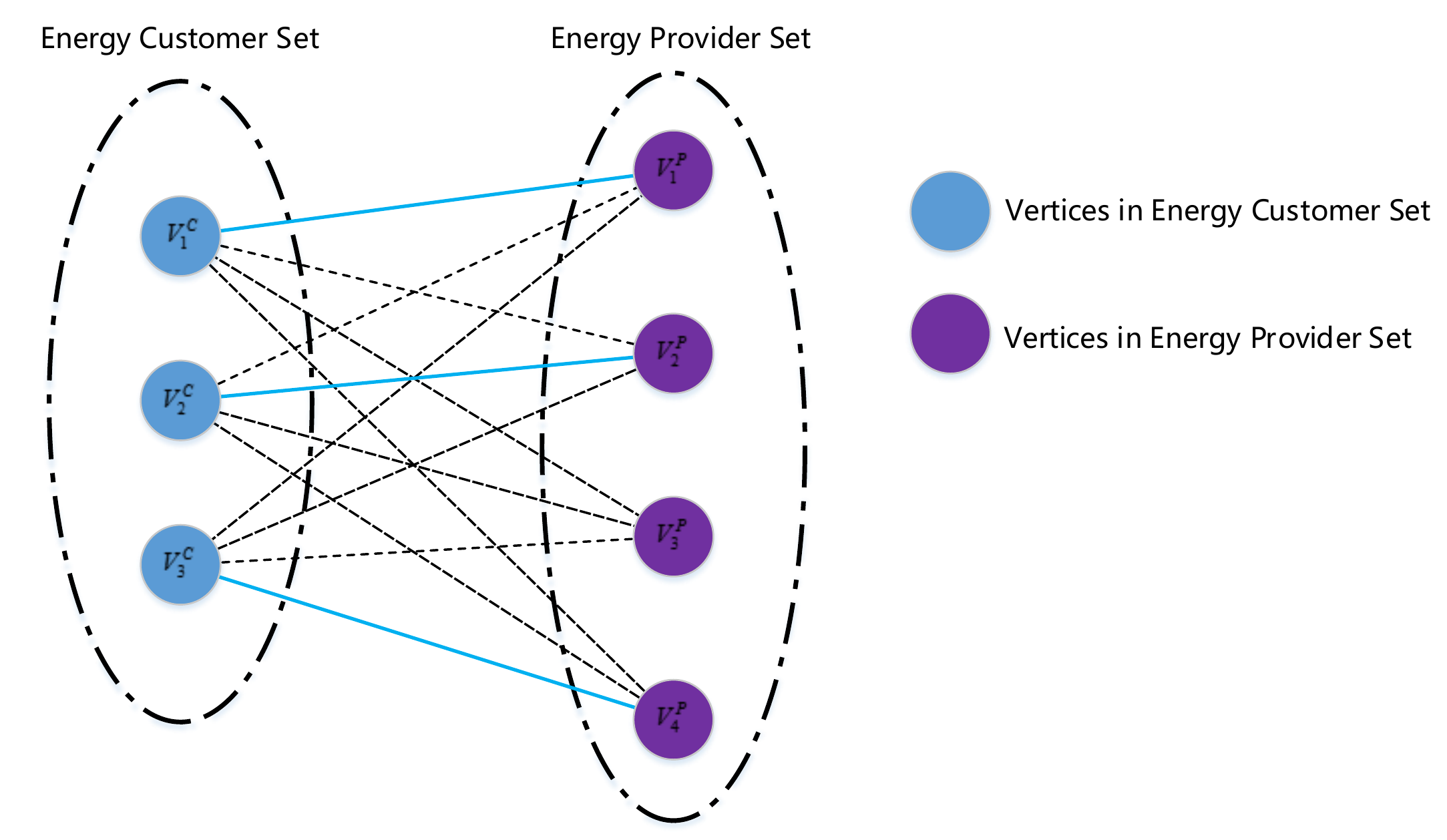}
\vspace{-5mm}
\caption{An illustration of the constructed bipartite graph.} \label{Fig-Bipartite_Graph}
\vspace{-5mm}
\end{figure}

In order to achieve efficient V2V matching between EVs as energy consumers and EVs as energy providers, we construct a bipartite graph $\mathcal{G} = \left(\mathcal{V},\mathcal{E}\right)$ according to the EV information collected in the data control center, as illustrated in Fig.~\ref{Fig-Bipartite_Graph}, where $\mathcal{V}$ denotes the vertex set and $\mathcal{E}$ denotes the edge set. The constructed bipartite graph is a complete, undirected, and weighted bipartite graph.

In the constructed bipartite graph, vertex set $\mathcal{V}$ is divided into two disjoint sets $\mathcal{V}^C$ and $\mathcal{V}^P$, where vertex $V_i^C$ in $\mathcal{V}^C$ denotes $\mbox{EV}_i^C$ as an energy consumer and vertex $V_j^P$ in $\mathcal{V}^P$ denotes $\mbox{EV}_j^P$ as an energy provider, $i\in\mathcal{N}$ and $j\in\mathcal{K}$.

In our proposed cooperative V2V charging based energy management protocol, through the V2V matching, our main target is to optimize the network social welfare, which is defined as the sum utility of all the involved EVs. Therefore, in the constructed weighted bipartite graph, the edge weight $W_{i,j}$ for edge $E = \langle V_i^C,V_j^P\rangle$ is defined as the sum utility of the two connected EVs, i.e.,
\begin{align} \label{Edge_Weight}
W_{i,j} = U_i^C \left(\mbox{EV}_j^{P}\right) + U_j^P \left(\mbox{EV}_i^{C}\right).
\end{align}

\textbf{Definition 1:} A \emph{matching} $\mathcal{M}$ of a bipartite graph $\mathcal{G}=\left(\mathcal{V},\mathcal{E}\right)$ is defined as a non-empty edge set that satisfies
\begin{enumerate}[1)]
  \item $\mathcal{M}$ is a subset of edge set $\mathcal{E}$, i.e., $\mathcal{M}\subseteq\mathcal{E}$;
  \item For any vertex $V\in\mathcal{V}$, at most one edge $E\in\mathcal{M}$ connects to $V$.
\end{enumerate}

Based on the constructed bipartite graph, in the following sections, we propose three effective and efficient V2V matching algorithms, i.e., the max-weight V2V matching algorithm, the EV-consumer-oriented stable V2V matching algorithm, and the EV-provider-oriented stable V2V matching algorithm.

\subsection{Max-Weight V2V Matching Algorithm}

In this section, based on the constructed bipartite graph, we propose a max-weight V2V matching algorithm in order to obtain the optimal matched EV pairs of energy consumers and energy providers, which can maximize the network social welfare (i.e., the sum edge weight of all the matched EV pairs). Therefore, the optimization problem can be transformed to the one aiming to find a maximum weight matching in the constructed weighted bipartite graph.

Before we propose the max-weight V2V matching algorithm, some definitions associated with the constructed weighted bipartite graph are demanded.

\textbf{Definition 2:} A \emph{perfect} matching is a matching $\mathcal{M}$ that every vertex is connected to some edge in $\mathcal{M}$. 

\textbf{Definition 3:} Vertex $V$ in a bipartite graph is \emph{matched} if it is an endpoint of an edge in a matching $\mathcal{M}$; otherwise, $V$ is \emph{free}.

\textbf{Definition 4:} A path is \emph{alternating} if its edges alternate between matching $\mathcal{M}$ and set $\mathcal{E}-\mathcal{M}$. An alternating path is \emph{augmenting} if both endpoints of the path are free.

\textbf{Definition 5:} A \emph{vertex labeling} $l$ in the bipartite graph $\mathcal{G}=\left(\mathcal{V},\mathcal{E}\right)$ is a function $l: \mathcal{V}\rightarrow\mathcal{R}$, where $\mathcal{R}$ is a real number set. A vertex labeling $l$ is \emph{feasible} such that $l\left(V_i^C\right) + l\left(V_j^P\right) \geq W_{i,j}$, $\forall~V_i^C\in\mathcal{V}^C$, $V_j^P\in\mathcal{V}^P$.

\textbf{Definition 6:} A weighted bipartite graph $\mathcal{G}=\left(\mathcal{V},\mathcal{E}_l\right)$ is an \emph{equality graph} based on a specific vertex labeling $l$, if $l\left(V_i^C\right) + l\left(V_j^P\right) = W_{i,j}$, $\forall~\langle V_i^C,V_j^P\rangle\in\mathcal{E}_l$, $V_i^C\in\mathcal{V}^C$, $V_j^P\in\mathcal{V}^P$.

Then, according to the Kuhn-Munkres theorem and Hungarian method \cite{HungarianMethod-MIT}, we design the max-weight V2V matching algorithm based on the constructed bipartite graph and the above definitions. The basic idea of the proposed max-weight V2V matching algorithm is to start with any feasible labeling $l$ and a matching $\mathcal{M}\subseteq\mathcal{E}_l$, and iteratively increase the size of $\mathcal{M}$ until $\mathcal{M}$ becomes a perfect matching. Note that in practical scenarios, the number of EVs as energy consumers may not be the same as that of EVs as energy providers in a specific matching period. In order to guarantee a perfect matching can be eventually found, during the construction of the bipartite graph, we can add some virtual EVs into the vertex set (either $\mathcal{V}^C$ or $\mathcal{V}^P$) with smaller number of EVs to make sure that both vertex sets have the same number of EVs, and set the edge weight of the edges connected to these virtual vertices to a sufficiently large negative value. For instance, if $N<K$, when constructing the bipartite graph, we will add $K-N$ virtual vertices into $\mathcal{V}^C$ and set $W_{i,j}=\mbox{Ne}$, $i = N+1,\cdots,K$, $j = 1,\cdots,K$, and $\mbox{Ne}$ is a large negative value. Hence, in the constructed bipartite graph, the number of the vertices in $\mathcal{V}^C$ is guaranteed to be equal to the number of the vertices in $\mathcal{V}^P$. For simplicity of descriptions, in the following, we assume that $N=K$ in the constructed bipartite graph.

The detailed procedure of the proposed max-weight V2V matching algorithm is given in Algorithm~\ref{Alg:Max-Weight-Matching}. In the proposed algorithm, given the constructed bipartite graph, we first initialize the vertex labeling as $l \left( V_i^C \right) = \mathop {\max} \limits_{V_j^P\in\mathcal{V}^{P}} W_{i,j}$ and $l \left( V_j^P \right) = 0$ that guarantees the initialized vertex labeling is feasible. Before stepping into the iteration process, we generate an initial matching $\mathcal{M}$ that comprises the edge with highest weight in the corresponding equality graph. In each round of the iteration process, we pick up a free vertex $V_i^C$ from the energy consumer set and try to add this selected vertex into the new matching by finding an augmenting path in which $V_i^C$ is one endpoint. If such an augmenting path $\mathcal{P}$ can be found in the current equality graph, we can update the matching $\mathcal{M}$ by $\mathcal{M} = \mathcal{M}\cup\mathcal{P} -  \mathcal{M}\cap\mathcal{P}$ and increase the size of the current matching $\mathcal{M}$ by $1$. Otherwise, we will update the vertex labeling as
\begin{align}\label{Vertex_Labeling_Update}
&\xi_l = \mathop {\min} \limits_{V_i^C\in\mathcal{S},V_j^P\notin\mathcal{T}} \left\{ l(V_i^C) + l(V_j^P) - W_{i,j} \right\} \nonumber \\
&l(V) = \left\{ \begin{array}{ll}
l(V) - \xi_l, \; &\mbox{if} \; V\in\mathcal{S} \nonumber \\
l(V) + \xi_l, \; &\mbox{if} \; V\in\mathcal{T} \nonumber \\
l(V), \; &\mbox{otherwise}
\end{array} \right.
\end{align}
in order to extend the current equality graph but meanwhile keep the vertex labeling still feasible. Note that every time we update the vertex labeling, the equality graph needs to be updated accordingly. The iteration process will terminate until each vertex in $\mathcal{V}^C$ and $\mathcal{V}^P$ is covered by $\mathcal{M}$, i.e., $\mathcal{M}$ becomes a perfect matching. The final output matching $\mathcal{M}$ indicates the matched EV pairs.

In addition, we have the following optimality property.

\textbf{Property 1:} Given $N = K$, the output perfect matching $\mathcal{M}$ of the proposed max-weight V2V matching algorithm is the optimal matching with maximum weight in the bipartite graph.
\begin{proof}
Let $\mathcal{M}'$ be any perfect matching in the constructed bipartite graph $\mathcal{G} = \left( \mathcal{V},\mathcal{E}\right)$ and $\mathcal{V}^{*}$ is a subset of $\mathcal{V}$.

Since every $V\in\mathcal{V}^{*}$ is covered exactly once by $\mathcal{M}'$, then we have
\begin{align}
W\left(\mathcal{M}'\right) &= \sum \limits_{E\in\mathcal{M'}} {W(E)} \nonumber \\
&\leq \sum \limits_{E\in\mathcal{M}'} {\left(l(V_i^C)+l(V_j^P)\right)} = \sum \limits_{V\in\mathcal{V}^{*}} {l(V)}
\end{align}
where $W\left(\mathcal{M}'\right)$ represents the sum edge weight of $\mathcal{M}'$.

According to the procedure of the proposed max-weight V2V matching algorithm, we have
\begin{align}
W\left(\mathcal{M}\right) = \sum \limits_{E\in\mathcal{M}} {W(E)} = \sum \limits_{V\in\mathcal{V}^{*}} {l(V)}
\end{align}
where $W\left(\mathcal{M}\right)$ represents the sum edge weight of $\mathcal{M}$.

Therefore, we can obtain that $W\left(\mathcal{M}'\right)\leq W\left(\mathcal{M}\right)$, which means that the output matching $\mathcal{M}$ is the optimal matching with maximum weight in the bipartite graph.
\end{proof}

In the proposed max-weight V2V matching algorithm, the matching size increases by $1$ through each phase, and thus there are at most $\mbox{max}(N,K)$ phases to reach a sub-perfect matching. Moreover, in each phase, at most $\mbox{max}(N,K)$ vertices can move from $\bar{\mathcal{S}}$ to $\mathcal{S}$ with vertex labeling re-calculation and updating time $\mathcal{O}(\mbox{max}(N,K))$ each time. Therefore, we can obtain the computational complexity of the proposed max-weight V2V matching algorithm as $\mathcal{O}\left(\mbox{max}(N,K)^3\right)$.

\begin{algorithm}\label{Alg:Max-Weight-Matching}
\caption{Max-Weight V2V Matching Algorithm}
\KwIn{Given the constructed bipartite graph $\mathcal{G}=\left(\mathcal{V},\mathcal{E}\right)$.}
\textbf{Initialization}: Initialize the feasible labeling $l$ as
\begin{align}
\left\{\begin{array}{ll}
l \left( V_i^C \right) = \mathop {\max} \limits_{V_j^P\in\mathcal{V}^{P}} W_{i,j}, \; V_i^C \in \mathcal{V}^{C} \nonumber \\
l \left( V_j^P \right) = 0, \; V_j^P \in \mathcal{V}^{P}
\end{array} \right.
\end{align}
\textbf{Step $1$}: Generate an initial matching $\mathcal{M}$ that comprises the edge with highest weight in equality graph $\mathcal{G}=\left(\mathcal{V},\mathcal{E}_l\right)$. \\
\textbf{Repeat} \\
\textbf{Step $2$}: Pick a free vertex $V_i^C$. \\
\textbf{Step $3$}: Initial set $\mathcal{S}$ and set $\mathcal{T}$ as $\mathcal{S}=\left\{V_i^C\right\}$ and $\mathcal{T}=\Phi$. \\
\textbf{Step $4$}: Define the neighbor set $\mathcal{N}_l(\mathcal{S})$ as
\begin{align}
\mathcal{N}_l (\mathcal{S}) = \left\{ V_j^P \in \mathcal{V}^{P} \mid V_i^C \in \mathcal{S}, \langle V_i^C,V_j^P \rangle \in \mathcal{E}_l \right\}. \nonumber
\end{align}
\textbf{Step $5$}: \If {$\mathcal{N}_l(\mathcal{S}) = \mathcal{T}$}{
Step $5.1$: Update the vertex labeling $l$ as
\begin{align}
&\xi_l = \mathop {\min} \limits_{V_i^C\in\mathcal{S},V_j^P\notin\mathcal{T}} \left\{ l(V_i^C) + l(V_j^P) - W_{i,j} \right\} \nonumber \\
&l(V) = \left\{ \begin{array}{ll}
l(V) - \xi_l, \; &\mbox{if} \; V\in\mathcal{S} \nonumber \\
l(V) + \xi_l, \; &\mbox{if} \; V\in\mathcal{T} \nonumber \\
l(V), \; &\mbox{otherwise}
\end{array} \right.
\end{align}
Step $5.2$: Update the corresponding equality graph and $\mathcal{N}_l(\mathcal{S})$.
}
\textbf{Step $6$}: \If {$\mathcal{N}_l(\mathcal{S}) \neq \mathcal{T}$}{
Step $6.1$: Find a free vertex in $\mathcal{N}_l(\mathcal{S}) - \mathcal{T}$. \\
Step $6.2$: \If {A free vertex $V_j^P$ can be found}{
Step $6.2.1$: Obtain an augmenting path $\mathcal{P} = V_i^C\rightarrow\cdots \rightarrow V_j^P$; \\
Step $6.2.2$: Update the matching $\mathcal{M}$ by $\mathcal{M} = \mathcal{M}\cup\mathcal{P} -  \mathcal{M}\cap\mathcal{P}$; \\
Step $6.2.3$: Go back to Step $2$.}
Step $6.3$: \Else{
Step $6.3.1$: Pick a vertex $V_j^P\in\mathcal{N}_l(\mathcal{S}) - \mathcal{T}$; \\
Step $6.3.2$: Obtain the vertex $Z$ that is matched to $V_j^P$; \\
Step $6.3.3$: Update $\mathcal{S} = \mathcal{S}\cup\{Z\}$ and $\mathcal{T} = \mathcal{T}\cup\{V_j^P\}$; \\
Step $6.3.4$: Go back to Step $5$.}
}
\textbf{Until} $\mathcal{M}$ becomes a perfect matching. \\
\textbf{Output} $\displaystyle \mathcal{M}$.
\end{algorithm}

\subsection{Stable V2V Matching Algorithm}

Although the proposed max-weight V2V matching algorithm can achieve an optimized network social welfare with low computational complexity, the obtained matching is not a stable one, since the proposed max-weight V2V matching algorithm is processed in a centralized manner and does not take each EV's individual rationality into consideration. However, in practical applications, each involved EV has its own preference for energy trading. Ignoring the EVs' individual rationality may lead to unstable and deviated behaviors in the energy trading market.

Therefore, in this section, we investigate the stable V2V matching by taking each involved EV's individual rationality into consideration and propose an optimal stable V2V matching algorithm with two-sided preferences, which can achieve optimized stable V2V matching also in terms of the network social welfare.

\subsubsection{Stable Matching}

\begin{figure}[!t]
\centering
\includegraphics[width=3.6in]{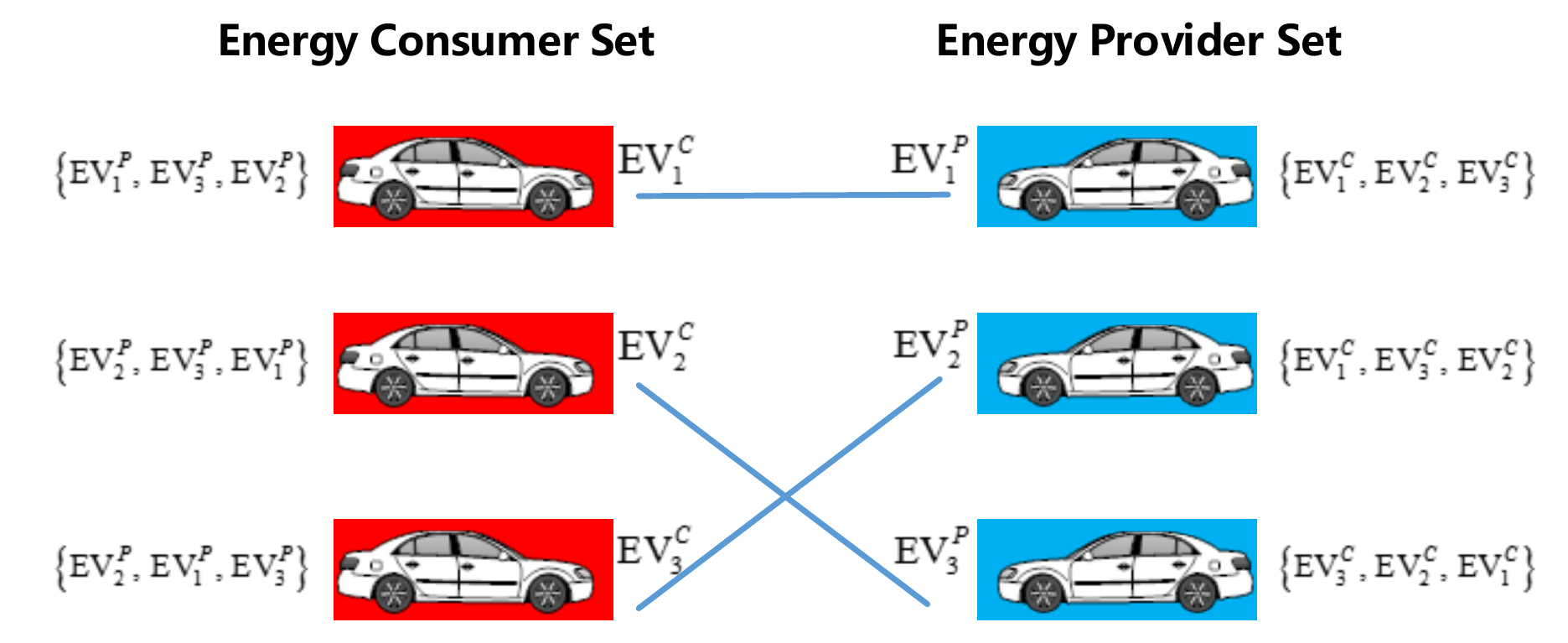}
\caption{A simple example of the one-to-one marriage model. The shown matching is a stable matching.} \label{Fig-Stable_Marriage}
\end{figure}

First, we introduce some basic concepts of the stable matching. As illustrated in Fig.~\ref{Fig-Stable_Marriage}, in our investigated matching problem, EVs as energy consumers and EVs as energy providers can be regarded as men and women in the one-to-one marriage model \cite{StableMarriage-XL-2011}, respectively. Each involved EV on one side (either energy consumer set or energy provider set) has a complete and transitive preference over the EVs on the other side, and can be represented by a rank order list including all the acceptable EVs on the other side. Note that if an EV (e.g., $\mbox{EV}_i^{C}$) prefers to remain single (i.e., unmatched) than being matched to another EV (e.g., $\mbox{EV}_j^{P}$), then $\mbox{EV}_j^{P}$ is said to be unacceptable to $\mbox{EV}_i^{C}$. We denote the preferences of $\mbox{EV}_i^{C}$ and $\mbox{EV}_j^{P}$ by $\mathcal{L}_i^{C}$ and $\mathcal{L}_j^{P}$, respectively, $i\in\mathcal{N}$, $j\in\mathcal{K}$.

In the investigated problem, each EV cares most about its own utility through the cooperative V2V charging. Therefore, based on the utilities of EVs as energy consumers and EVs as energy providers, we define the prefer relation for $\mbox{EV}_i^{C}$ and $\mbox{EV}_j^{P}$ (denoted by $\succ_{\mbox{EV}_i^{C}}$ and $\succ_{\mbox{EV}_j^{P}}$) in Definition 7 and Definition 8, respectively.

\textbf{Definition 7:} For $\mbox{EV}_i^{C}$, it prefers $\mbox{EV}_j^{P}$ to $\mbox{EV}_{j'}^{P}$, i.e., $\mbox{EV}_j^{P} \succ_{\mbox{EV}_i^{C}} \mbox{EV}_{j'}^{P}$, if $U_i^{C}\left(\mbox{EV}_j^{P}\right) > U_i^{C}\left(\mbox{EV}_{j'}^{P}\right)$, $i\in\mathcal{N}$, $j,{j'}\in\mathcal{K}$, $j\neq{j'}$.

\textbf{Definition 8:} For $\mbox{EV}_j^{P}$, it prefers $\mbox{EV}_i^{C}$ to $\mbox{EV}_{i'}^{C}$, i.e., $\mbox{EV}_i^{C} \succ_{\mbox{EV}_j^{P}} \mbox{EV}_{i'}^{C}$, if $U_j^{P}\left(\mbox{EV}_i^{C}\right) > U_j^{P}\left(\mbox{EV}_{i'}^{C}\right)$, $j\in\mathcal{K}$, $i,{i'}\in\mathcal{N}$, $i\neq{i'}$.

In order to judge whether a matching is a stable one, we need to introduce the following definitions first.

\textbf{Definition 9:} A matching $\mathcal{M}$ is \emph{individual rational} to all the EVs, if and only if there does not exist an involved EV that prefers being unmatched to being matched within $\mathcal{M}$.

\textbf{Definition 10:} A matching $\mathcal{M}$ is \emph{blocked} by a pair of EVs if they prefer each other than the paired EVs through the matching $\mathcal{M}$. Such a pair is called a \emph{blocking set} in general.

Note that if there is a blocking set in the matching, the EVs involved will have an incentive to break up and form new marriages. Therefore, the matching is considered to be unstable.

\textbf{Definition 11:} A matching $\mathcal{M}$ is \emph{stable} if and only if $\mathcal{M}$ is individual rational and is not blocked by any pair of EVs.

According to the above stable matching definitions, we have that the matching shown in Fig.~\ref{Fig-Stable_Marriage} is a stable matching for the given preferences of the EVs. It is proved by the classical deferred acceptance algorithm \cite{DA-Algorithm-2008} that at least one stable matching exists for every stable marriage problem. Then, the problem comes to how find a stable matching effectively and efficiently.

\subsubsection{EV-Consumer-Oriented and EV-Provider-Oriented V2V Matching Algorithms}

The Gale-Shapley algorithm \cite{GS-Algorithm-1962} has been proposed as an efficient method to find a stable one-to-one matching between men and women in the stable marriage problem. Similarly, in our investigated problem, EVs as energy consumers and EVs as energy providers can be regarded as men and women, respectively. Then, referring to the Gale-Shapley algorithm, we design the EV-consumer-oriented and EV-provider-oriented V2V matching algorithms in order to obtain the stable matching between EVs as energy consumers and EVs as energy providers.

The detailed procedure of the EV-consumer-oriented V2V matching algorithm is given in Algorithm~\ref{Alg:Gale-Shapley-Consumer}. In the proposed EV-consumer-oriented V2V matching algorithm, at first each EV as energy consumer proposes to its most preferred EV as energy provider according to its preference list. For each EV as energy provider, if more than one acceptable proposals are received, it will hold the most preferred one and reject all the others. In each following round, any EV as energy consumer that is rejected in the previous round updates its preference list by deleting the first element and makes a new proposal to its current most preferred partner who has not yet rejected it, or makes no proposals if no acceptable choices remains. Each EV as energy provider holds its most preferred offer up to the current round and rejects all the rest. The iteration process will terminate until no further proposals can be made, that is, there is either no EVs as energy consumers still rejected or no remaining acceptable choices in the preference lists of EVs as energy consumers being rejected.

The proposed EV-consumer-oriented V2V matching algorithm can be easily transformed into an EV-provider-oriented one by swapping the roles of EVs as energy consumers and EVs as energy providers, that is, EVs as energy providers make proposals to EVs as energy consumers and EVs as energy consumers decide to hold or reject the received proposals, as given in Algorithm~\ref{Alg:Gale-Shapley-Provider}.

According to the procedure of the proposed EV-consumer-oriented and EV-provider-oriented V2V matching algorithms, the EVs make proposals or decide to hold or reject the received proposals in an independent and distributed manner, therefore, the computational complexity of both the proposed EV-consumer-oriented and EV-provider-oriented V2V matching algorithms is $\mathcal{O}(N+K)$.


Note that although the provided EV-consumer-oriented and EV-provider-oriented V2V matching algorithms can both realize stable matchings, they have significant consequences. The EV-consumer-oriented algorithm yields an EV-consumer-optimal stable matching, in which each EV as energy consumer has the best matched partner that it can have in any stable matching, whereas the EV-provider-oriented algorithm leads to an EV-consumer-optimal output. This property is referred to as the polarization of stable matchings \cite{GS-Algorithm-1962}.


\begin{algorithm}\label{Alg:Gale-Shapley-Consumer}
\caption{EV-Consumer-Oriented V2V Matching Algorithm}
\KwIn{Given the constructed bipartite graph $\mathcal{G}=\left(\mathcal{V},\mathcal{E}\right)$.}
\textbf{Step $1$}: Set up the preference lists of vertices $V_i^{C}$ and $V_j^{P}$ denoted by $\mathcal{L}_i^{C}$ and $\mathcal{L}_j^{P}$, respectively. \\
\textbf{Step $2$}: Initialize $\mathcal{U}$ including all the unmatched EVs as energy consumers, i.e., $\mathcal{U}=\left\{V_i^{C}\mid i\in\mathcal{N} \right\}$. \\
\textbf{Step $3$}: Initialize $\mathcal{H}_j = \Phi$ as the current hold of $V_j^{P}$, $j\in\mathcal{K}$. \\
\textbf{Repeat} \\
\textbf{Step $4$}: $V_i^{C}$ proposes to the vertex that locates first in its preference list $\mathcal{L}_i^{C}$, $\forall V_i^C\in\mathcal{U}$. \\
\textbf{Step $5$}: \For {$j = 1, 2, \cdots, K$}{
\If {$V_j^{P}$ receives a more preferred proposal from $V_{i'}^{C}$ than the current hold}{
$V_{i'}^{C}$ is removed from $\mathcal{U}$ and the current hold $\mathcal{H}_j$ is added into $\mathcal{U}$; \\
$V_j^{P}$ updates $\mathcal{H}_j = V_{i'}^{C}$.}
\Else {$V_j^{P}$ rejects the received proposals and continues the current hold.}
}
\textbf{Step $6$}: \For {$i = 1, 2, \cdots, N$}{
\If{$V_i^C$ is rejected in this round}{Update the preference list $\mathcal{L}_i^C$ by deleting the first element in $\mathcal{L}_i^C$.}}
\textbf{Until} $\mathcal{U}$ is empty or each vertex in $\mathcal{U}$ has an empty preference list. \\
\textbf{Step $7$}: \For {$j = 1, 2, \cdots, K$}{
\If {$\mathcal{H}_j \neq \Phi$}{
Add $V_j^P$ and $\mathcal{H}_j$ as a matched pair into $\mathcal{M}$.}
}
\textbf{Output} $\displaystyle \mathcal{M}$.
\end{algorithm}

\begin{algorithm}\label{Alg:Gale-Shapley-Provider}
\caption{EV-Provider-Oriented V2V Matching Algorithm}
\KwIn{Given the constructed bipartite graph $\mathcal{G}=\left(\mathcal{V},\mathcal{E}\right)$.}
\textbf{Step $1$}: Set up the preference lists of vertices $V_i^{C}$ and $V_j^{P}$ denoted by $\mathcal{L}_i^{C}$ and $\mathcal{L}_j^{P}$, respectively. \\
\textbf{Step $2$}: Initialize $\mathcal{U}$ including all the unmatched EVs as energy providers, i.e., $\mathcal{U}=\left\{V_j^{P} \mid j\in\mathcal{K} \right\}$. \\
\textbf{Step $3$}: Initialize $\mathcal{H}_i = \Phi$ as the current hold of $V_i^{C}$, $i\in\mathcal{N}$. \\
\textbf{Repeat} \\
\textbf{Step $4$}: $V_j^{P}$ proposes to the vertex that locates first in its preference list $\mathcal{L}_j^{P}$, $\forall V_j^{P}\in\mathcal{U}$. \\
\textbf{Step $5$}: \For {$i = 1, 2, \cdots, N$}{
\If {$V_i^{C}$ receives a more preferred proposal from $V_{j'}^{P}$ than the current hold}{
$V_{j'}^{P}$ is removed from $\mathcal{U}$ and the current hold $\mathcal{H}_i$ is added into $\mathcal{U}$; \\
$V_i^{C}$ updates $\mathcal{H}_i = V_{j'}^{P}$.}
\Else {$V_i^{C}$ rejects the received proposals and continues the current hold.}
}
\textbf{Step $6$}: \For {$j = 1, 2, \cdots, K$}{
\If{$V_j^P$ is rejected in this round}{Update the preference list $\mathcal{L}_j^P$ by deleting the first element in $\mathcal{L}_j^P$.}}
\textbf{Until} $\mathcal{U}$ is empty or each vertex in $\mathcal{U}$ has an empty preference list. \\
\textbf{Step $6$}: \For {$i = 1, 2, \cdots, N$}{
\If {$\mathcal{H}_i \neq \Phi$}{
Add $V_i^C$ and $\mathcal{H}_i$ as a matched pair into $\mathcal{M}$.}
}
\textbf{Output} $\displaystyle \mathcal{M}$.
\end{algorithm}

\section{Simulations and Discussions}%

To evaluate the efficiency of the proposed cooperative V2V charging based energy management protocol for the EVs in the investigated system, we conduct the following simulations. As a performance comparison baseline, we employ the traditional EV charging protocol, in which the EVs as energy consumers choose to get charged at the nearest charging station. The utilities of EVs as energy consumers when getting charged at the selected charging station are given in (\ref{Utility_EV_C_CS}).

\subsection{Simulation Parameters}
In the simulations, we consider a $20~\mbox{km}\times20~\mbox{km}$ urban network with $50$ EVs driving in it. The EVs are initialized at random locations with a random driving direction and we assume that in a specific energy trading task period, the EVs follow uniform rectilinear motion. There are two charging stations located at $(10~\mbox{km},5~\mbox{km})$ and $(10~\mbox{km},15~\mbox{km})$, respectively, and $25$ available parking lots located uniformly in the simulated scenario. We randomly select $N$ EVs as energy consumers that demand power for further driving towards their individual destinations and $K$ EVs as energy providers that have surplus power for energy trading. The detailed simulation parameters are listed in Table~\ref{Simulation_Parameters}.


\begin{table*}[!t]
\renewcommand{\arraystretch}{1.3}
\caption{Simulation Parameters} \label{Simulation_Parameters}
\centering
\begin{tabular}{|l|l|}
\hline Parameters & Value \\ \hline
{EV's Velocity $v_i^C$ and $v_j^P$} & {Uniform Distribution Between $20$ and $60~\mbox{km/h}$} \\
Unit Power Trading Price $p_t$ & {$15$ c/kWh \cite{Book-EV-2016}}\\
Electricity Selling Price $p_s$ & {$18$ c/kWh \cite{Book-EV-2016}}\\
The V2V Power Transfer Efficiency $\eta$ & {$0.95$ \cite{V2X-Liu-2013}}  \\
The Moving Energy Cost for $\mbox{EV}_i^{C}$ $\beta_i^C$ & {Uniform Distribution Between $0.2$ and $0.5~\mbox{kWh/km}$} \\
The Moving Energy Cost for $\mbox{EV}_j^{P}$ $\beta_j^P$ & {Uniform Distribution Between $0.2$ and $0.5~\mbox{kWh/km}$} \\
The Required Power Amount $a_i^C$ & {Uniform Distribution Between $20$ and $40~\mbox{kWh}$} \\
Number of EVs as Energy Consumers $N$ & [10, 15, 20, 25, 30, 35, 40] \\
Number of EVs as Energy Providers $K$ & [10, 15, 20, 25, 30, 35, 40] \\
\hline
\end{tabular}
\end{table*}

\subsection{EV Utility Comparison}

\begin{figure}[!t]
\centering
\includegraphics[width=3.6in]{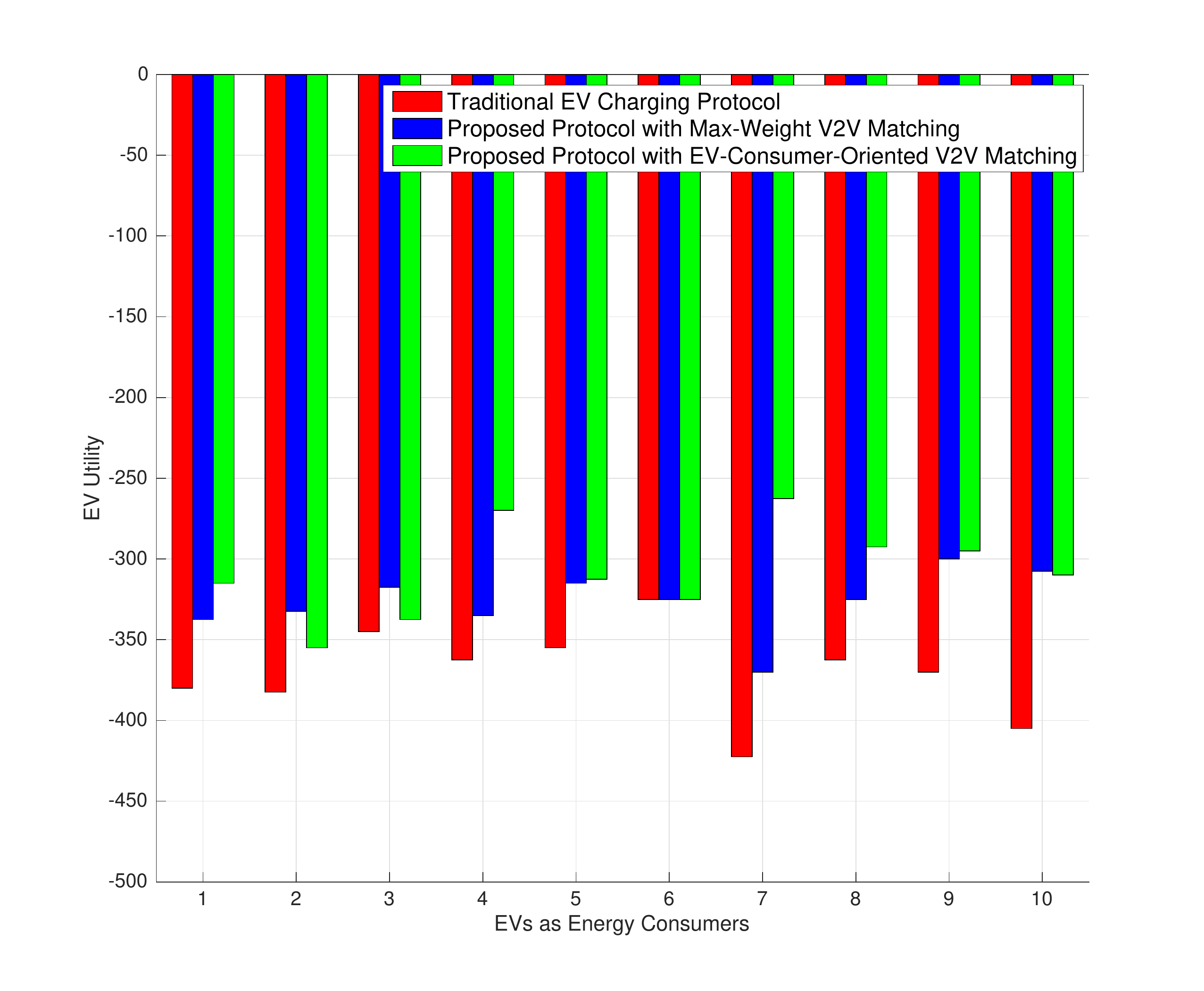}
\caption{Utility performance comparison for EVs as energy consumers ($N=K=10$).} \label{Fig-Utility_Comparison_Consumer}
\end{figure}

\begin{figure}[!t]
\centering
\includegraphics[width=3.6in]{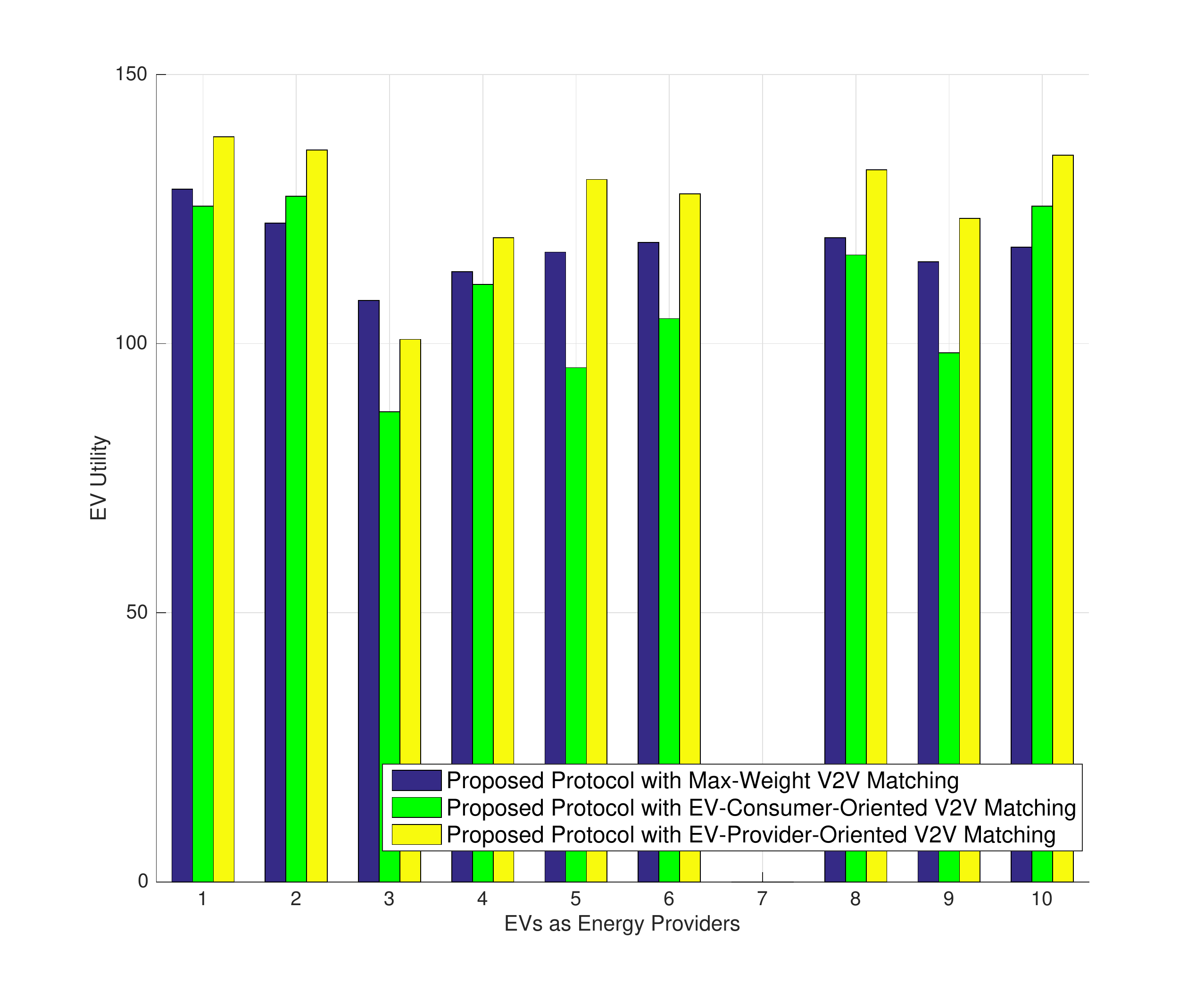}
\caption{Utility performance comparison for EVs as energy providers ($N=K=10$).}\label{Fig-Utility_Comparison_Provider}
\end{figure}

In Fig.~\ref{Fig-Utility_Comparison_Consumer}, we simulate the utilities of EVs as energy consumers with the traditional EV charging protocol and the proposed cooperative V2V charging based energy management protocol. Here in our proposed energy management protocol, the max-weight V2V matching algorithm and the EV-consumer-oriented stable V2V matching algorithm are selected to obtain the V2V matching solutions. From Fig.~\ref{Fig-Utility_Comparison_Consumer}, we can clearly find that with our proposed energy management protocol, the utilities of EVs as energy consumers are improved significantly, leading to smarter and more effective charging behaviors. It can be also found that $\mbox{EV}_6^C$ has the same utility value with our proposed energy management protocol and the traditional EV charging protocol. This implies that for $\mbox{EV}_6^C$, cooperative V2V charging with other EVs as energy providers in the network cannot lead to a better utility than to get charged at the charging station. Therefore, it finally chooses to get charged at the nearest charging station based on the feedback decisions from the data control center.

In Fig.~\ref{Fig-Utility_Comparison_Provider}, we simulate the utilities of EVs as energy providers with the proposed cooperative V2V charging based energy management protocol. All the proposed V2V matching algorithms are selected for comparison. From Fig.~\ref{Fig-Utility_Comparison_Provider}, we can see most EVs as energy providers can achieve a positive utility value, which makes the EVs that have extra power have an incentive to participate in the cooperative V2V charging process as energy providers. It can be found that $\mbox{EV}_7^{P}$ has zero utility value, which means $\mbox{EV}_7^P$ doesn't find an effective partner for energy trading (i.e., unmatched) with our proposed energy management protocol. There are two reasons resulting in this situation. First, $\mbox{EV}_7^P$ cannot achieve a positive utility based on the V2V matching solutions and thus it prefers to be unmatched. Second, the matched partner of $\mbox{EV}_7^P$ cannot achieve a better utility through cooperative V2V charging than to get charged at the charging station (e.g., $\mbox{EV}_6^C$ in Fig.~\ref{Fig-Utility_Comparison_Consumer}), and thus the matched partner leaves the matched relation, making $\mbox{EV}_7^P$ also become unmatched. We can also find the utilities of EVs as energy providers with the EV-provider-oriented stable V2V matching algorithm are never smaller than those with the EV-consumer-oriented stable V2V matching algorithm. This is because the EV-provider-oriented stable V2V matching algorithm can always yield the EV-provider-optimal output. But we should note that the EV-consumer-optimal and EV-provider-optimal properties are only guaranteed among all the possible stable matchings not all the matchings. Hence, there is still a possibility that the utilities of some EVs as energy consumers/energy providers achieved by the proposed max-weight V2V matching algorithm are larger than those with the proposed EV-consumer-oriented/EV-provider-oriented V2V matching algorithm, as illustrated in Figs.~\ref{Fig-Utility_Comparison_Consumer} and \ref{Fig-Utility_Comparison_Provider}.

In the simulations, based on the individual and location information of the randomly initialized EVs, the preference list of each EV is obtained according to the prefer relation definitions. The relative location topology between EVs as energy consumers and EVs as energy providers is the key factor to determine the preference of the EVs, since either an EV as energy consumer or an EV as energy provider prefers to achieve energy trading with a nearby EV at the nearest parking lot to reduce its energy cost and time cost. According to the simulation results, the preference lists of EVs as energy consumers play a more important role in the EV-consumer-oriented V2V matching algorithm, leading to the EV-consumer-optimal V2V stable matching solutions, where the preference of EVs as energy consumers has a higher priority to be satisfied. Whereas in the EV-provider-oriented V2V matching algorithm, the preference of EVs as energy providers is satisfied in a priority order, resulting in the EV-provider-optimal V2V stable matching solutions.

\subsection{Network Social Welfare Comparison}


\begin{figure}[!t]
\centering
\subfigure[$N=10$, $K$ changes] {\includegraphics[height=3.0in,width=3.6in]{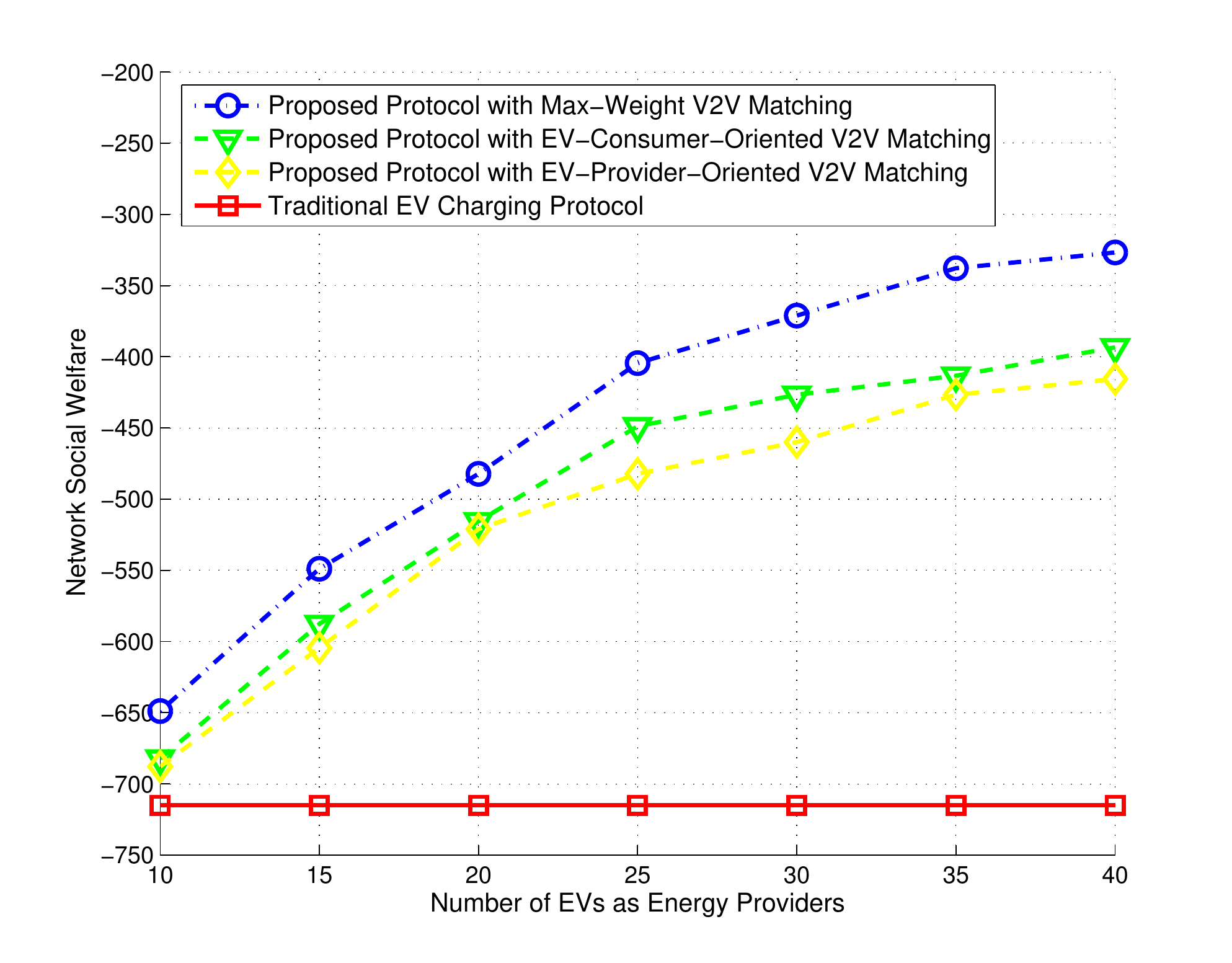}}
\subfigure[$K=10$, $N$ changes] {\includegraphics[height=3.0in,width=3.6in]{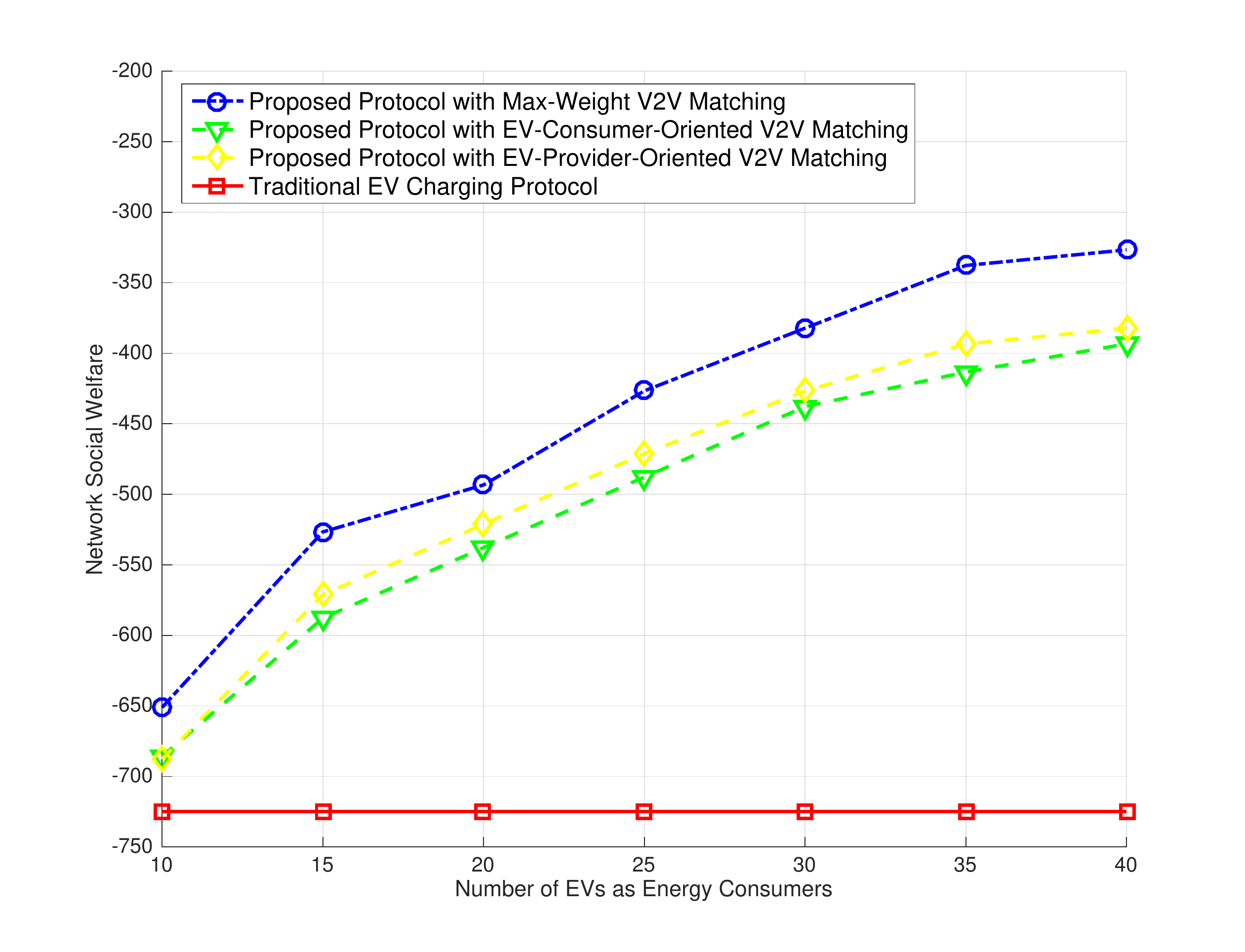}}
\caption{Network social welfare performance comparison with different number of EVs.}
\label{Fig-Social_Welfare}
\end{figure}

In Figs.~\ref{Fig-Social_Welfare}a and \ref{Fig-Social_Welfare}b, we compare the network social welfare performance of the traditional EV charging protocol and the proposed cooperative V2V charging based energy management protocol with different V2V matching algorithms. For comparison fairness, the network social welfare with the traditional EV charging protocol is defined as the sum utility of the EVs as energy consumers and the charging stations, where the utility of the charging stations is given by $U_{CS} = \sum \limits_{i\in\mathcal{N}} {p_s a_i^C}$. Note that the obtained network social welfare is an average over 10000 simulations with randomly initialized EVs in the simulated scenario, and the standard deviation from the average is small. 
From Figs.~\ref{Fig-Social_Welfare}a and \ref{Fig-Social_Welfare}b, we can see that our proposed energy management protocol can achieve an obvious improvement compared with the traditional EV charging protocol in terms of network social welfare. For the three proposed V2V matching algorithms, the max-weight V2V matching algorithm can always achieve the highest network social welfare, but the obtained matching does not take the individual rationality into consideration and thus is an unstable matching. We can also find that when the number of EVs as energy providers is larger than that of EVs as energy consumers, the EV-consumer-oriented V2V matching algorithm achieves better network performance compared with the EV-provider-oriented one. This is because in such a case, EVs as energy consumers have more available energy trading candidates (i.e., EVs as energy providers) to select, and the EV-consumer-oriented V2V matching algorithm always leads to the EV-consumer-optimal stable V2V matching, which can more sufficiently utilize the opportunistic selection gain on the side of EVs as energy consumers to obtain better network performance. Similarly, when the number of EVs as energy providers is smaller than that of EVs as energy consumers, the EV-provider-oriented V2V matching algorithm achieves better network performance compared with the EV-consumer-oriented one.

\subsection{Energy Consumption Reduction}


\begin{figure}[!t]
\centering
\subfigure[$N=10$, $K$ changes] {\includegraphics[height=3.0in,width=3.6in]{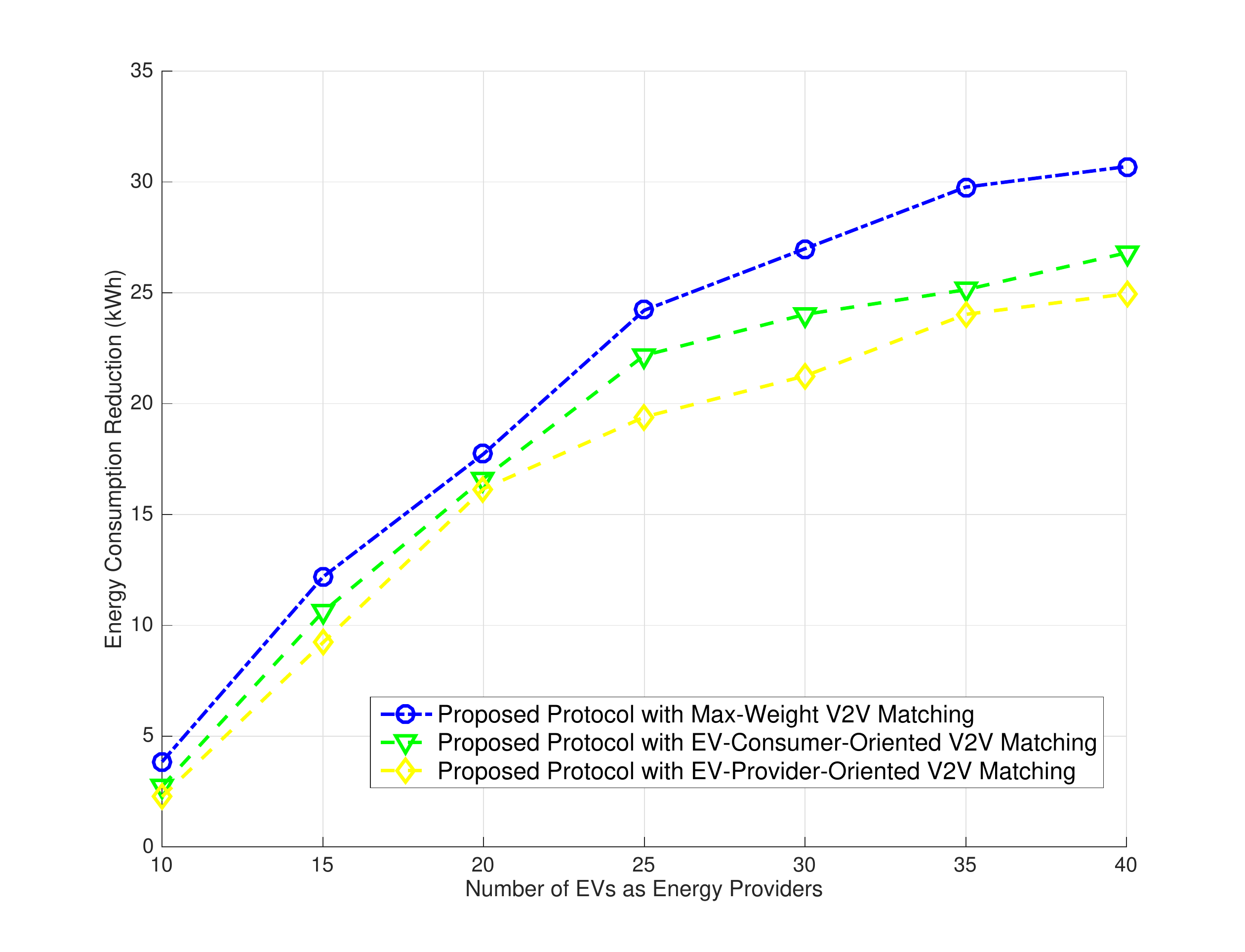}}
\subfigure[$K=10$, $N$ changes] {\includegraphics[height=3.0in,width=3.6in]{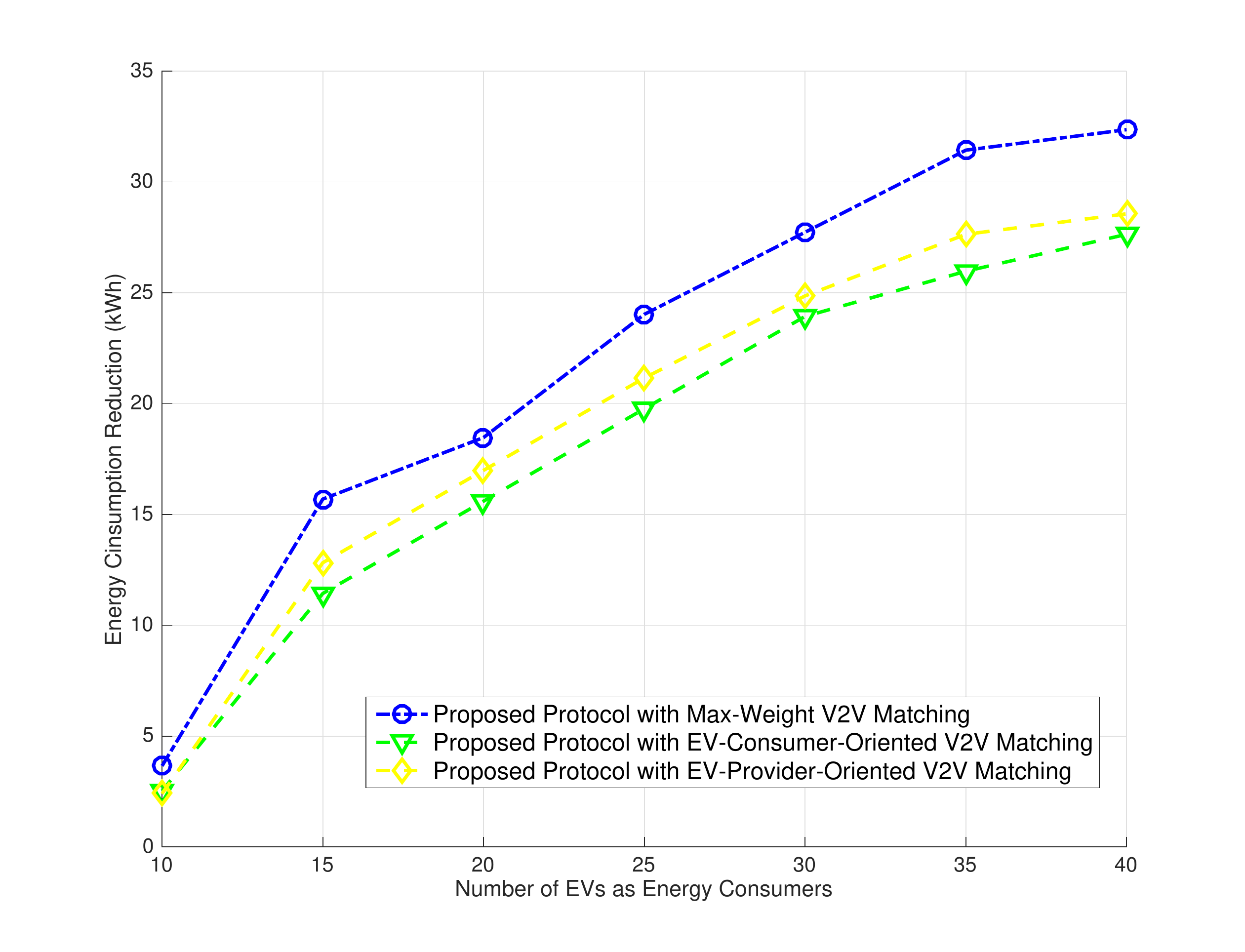}}
\caption{Energy consumption reduction through the proposed energy management protocol with different V2V matching algorithms.}
\label{Fig-Energy_Reduction}
\end{figure}

In Figs.~\ref{Fig-Energy_Reduction}a and \ref{Fig-Energy_Reduction}b, we simulate the energy consumption reduction of all the involved EVs through the proposed cooperative V2V charging based energy management protocol with different V2V matching algorithms, compared with the traditional EV charging protocol where all the EVs as energy consumers choose to get charged at the nearest charging station. Actually, the energy consumption reduction is calculated as the network energy cost (i.e., the sum of energy cost of EVs as energy consumers and EVs as energy providers that finally participate in energy trading) difference between our proposed energy management protocol and the traditional EV charging protocol. From Figs.~\ref{Fig-Energy_Reduction}a and \ref{Fig-Energy_Reduction}b, we can clearly find that the energy consumption of the involved EVs can be reduced effectively through our proposed energy management protocol with all the three V2V matching algorithms. This leads to a more flexible and smarter energy management for the EV system. The simulated energy consumption reduction is subject to a limited simulated scenario, but we can envision that in the fast-developing EV systems here and there, with our proposed cooperative V2V charging based energy management protocol, a huge amount of energy can be saved day by day.

\subsection{Computation Time}

\begin{figure}[!t]
\centering
\includegraphics[width=3.6in]{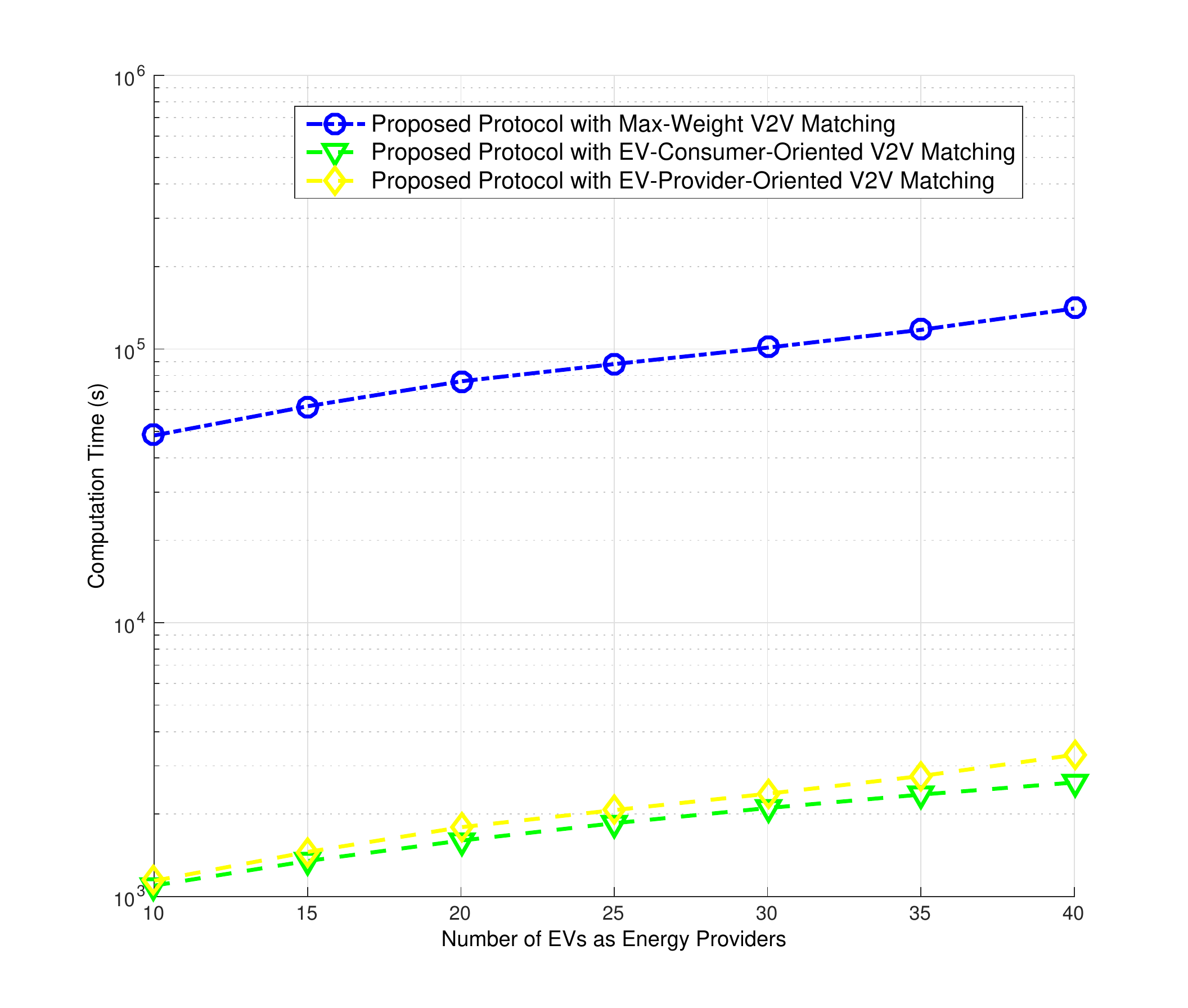}
\caption{Computation time comparison of the proposed energy management protocol with different V2V matching algorithms.} \label{Fig-Computation_Time}
\end{figure}

In Fig.~\ref{Fig-Computation_Time}, we compare the computation time of the proposed cooperative V2V charging based energy management protocol with different V2V matching algorithms. Note that here the computation time is calculated based on $1000$ times realizations with randomly initialized EVs in the simulated scenario and the y axis in Fig.~\ref{Fig-Computation_Time} is set as a $\log$-axis due to the large computation time difference between the max-weight V2V matching algorithm and the other two V2V matching algorithms. From Fig.~\ref{Fig-Computation_Time}, we can see that the max-weight V2V matching algorithm has significantly higher computation time compared with the EV-consumer-oriented and EV-provider-oriented V2V matching algorithms, which can be regarded as a cost to achieve the optimal V2V matching in terms of network social welfare. The EV-consumer-oriented and EV-provider-oriented V2V matching algorithms have similar computation time. The obtained computation time results agree with our theoretical analysis for the computational complexity of the designed V2V matching algorithms in Subsections IV-B and IV-C.

\section{Conclusions and Future Work}%

In this paper, we investigated flexible energy management through active power transfer cooperation between EVs as energy consumers and EVs as energy providers in an energy Internet based EV system. We first introduced a developed cooperative V2V charging concept. Then, we proposed a cooperative V2V charging based energy management protocol with different V2V matching algorithms, which can help the EVs achieve flexible and smart charging/discharging behaviors. Simulation results indicated that our proposed cooperative V2V charging based energy management protocol with different V2V matching algorithms can effectively improve the utilities of the EVs and reduce the network energy consumption. The max-weight V2V matching algorithm can always lead to best network performance but it does not take the individual rationality of the EVs into consideration, which means the obtained V2V matchings may be not stable if the EVs can make decisions based on their own utilities. Whereas the proposed EV-consumer-oriented and EV-provider-oriented V2V matching algorithms can output EV-consumer-optimal and EV-provider-optimal stable V2V matchings with very low computational complexity, respectively. 

In our future work, we will extend the investigated scenario to a more general energy Internet based system, where nearby charging stations, any available smart houses with distributed energy resources (DERs), and the EVs can be all involved in flexible energy management in a cooperative manner. This will result in more flexible and smarter charging/discharging behaviors of EVs. Since the charging stations and smart houses can be open to multiple EVs as energy consumers simultaneously, we intend to employ many-to-many matching model to formulate such an energy management problem.


\begin{IEEEbiography}[{\includegraphics[width=1in,height=1.25in,clip,keepaspectratio]{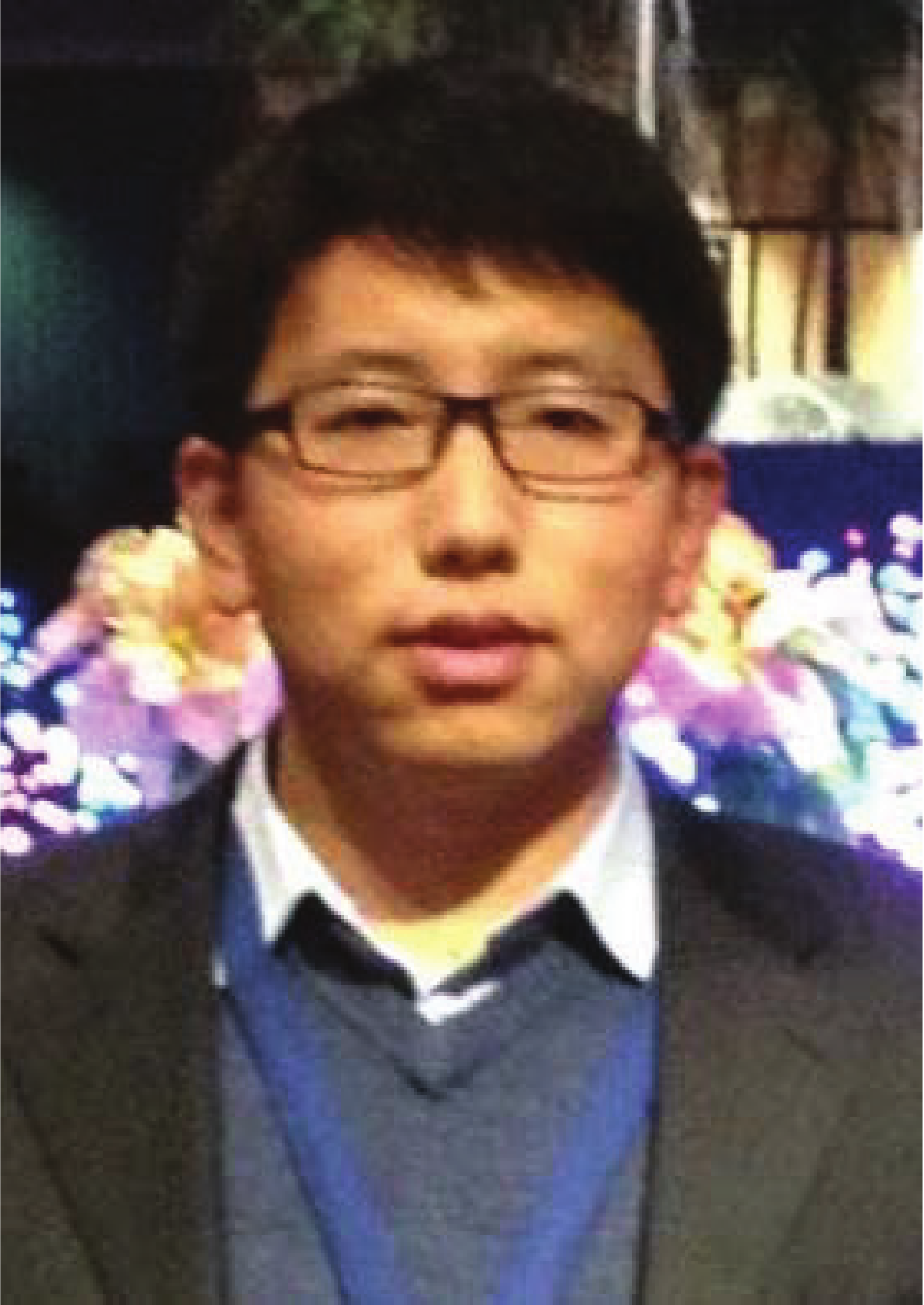}}]{Rongqing Zhang}
(S'11-M'15) received the B.S. and Ph.D. degrees from Peking University, Beijing, China, in 2009 and 2014, respectively. Since 2014, he has been a post-doctoral research fellow at Colorado State University, CO, USA. He has published two book chapters and more than 60 papers in refereed journals and conference proceedings. His current research interests include physical layer security, vehicular communications and networking, and electric vehicles.

Dr. Zhang was the recipient of the Academic Award for Excellent Doctoral Students, Ministry of Education of China, the co-recipient of the First-Class Natural Science Award, Ministry of Education of China, and received the Best Paper Awards at IEEE ITST 2012 and ICC 2016. He was also awarded as International Presidential Fellow of Colorado State University in 2017.
\end{IEEEbiography}

\begin{IEEEbiography}[{\includegraphics[width=1in,height=1.25in,clip,keepaspectratio]{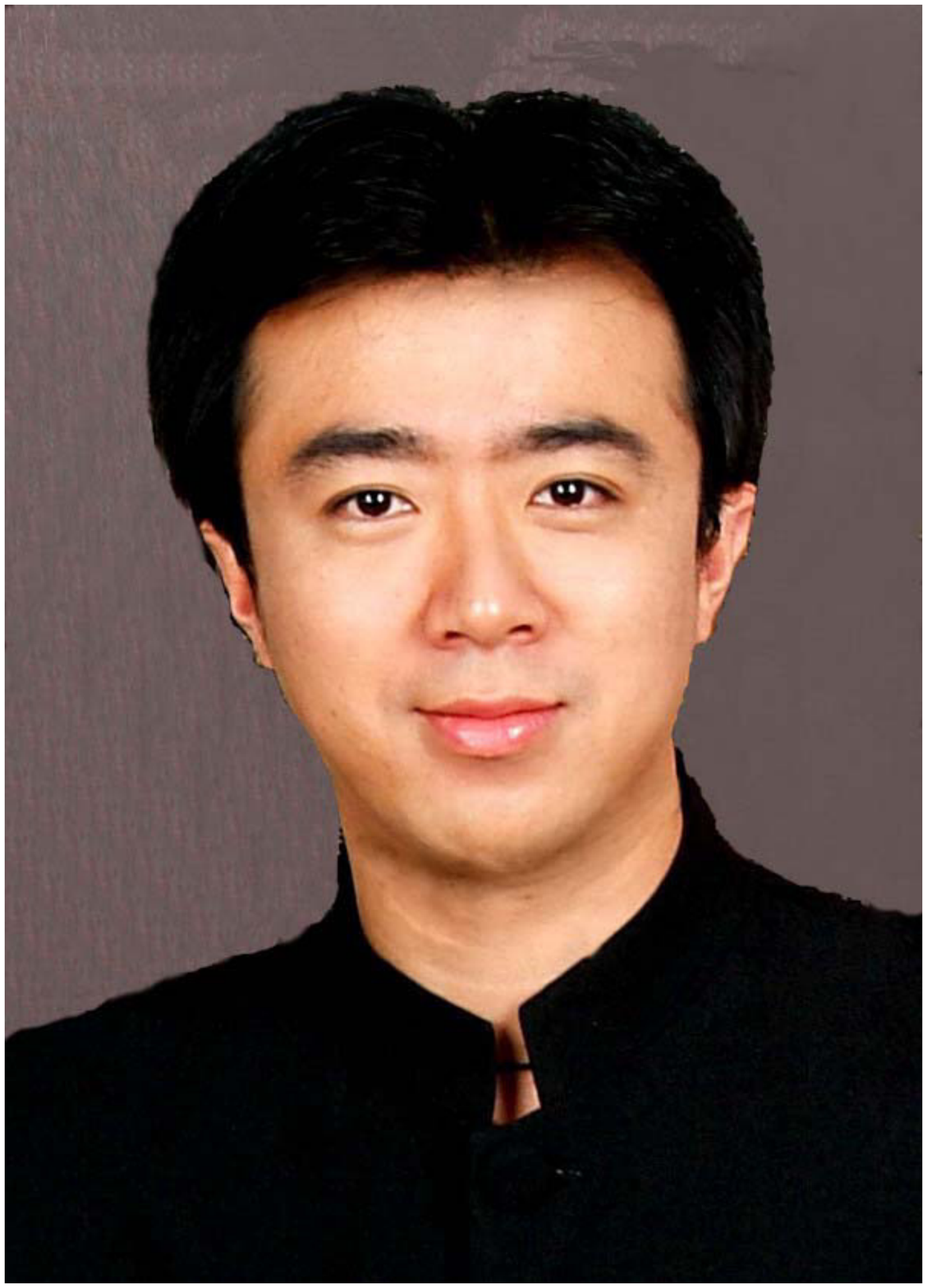}}]{Xiang Cheng}
(S'05-M'10-SM'13) received the Ph.D. degree from Heriot-Watt University and the University of Edinburgh, Edinburgh, U.K., in 2009, where he received the Postgraduate Research Thesis Prize. He is currently a Professor at Peking University. His general research interests are in areas of channel modeling and mobile communications, subject on which he has published more than 160 journal and conference papers, 3 books and 6 patents. 

Dr. Cheng was the recipient of the IEEE Asia Pacific (AP) Outstanding Young Researcher Award in 2015, the co-recipient for the 2016 IEEE JSAC Best Paper Award: Leonard G. Abraham Prize, the NSFC Outstanding Young Investigator Award, the Second-Rank Award in Natural Science, Ministry of Education in China, and received the Best Paper Awards at IEEE ITST'12, ICCC'13, ITSC'14, ICC'16, and ICNC'17. He has served as Symposium Leading-Chair, Co-Chair, and a Member of the Technical Program Committee for several international conferences. He is now an Associate Editor for IEEE Transactions on Intelligent Transportation Systems.
\end{IEEEbiography}

\begin{IEEEbiography}[{\includegraphics[width=1in,height=1.25in,clip,keepaspectratio]{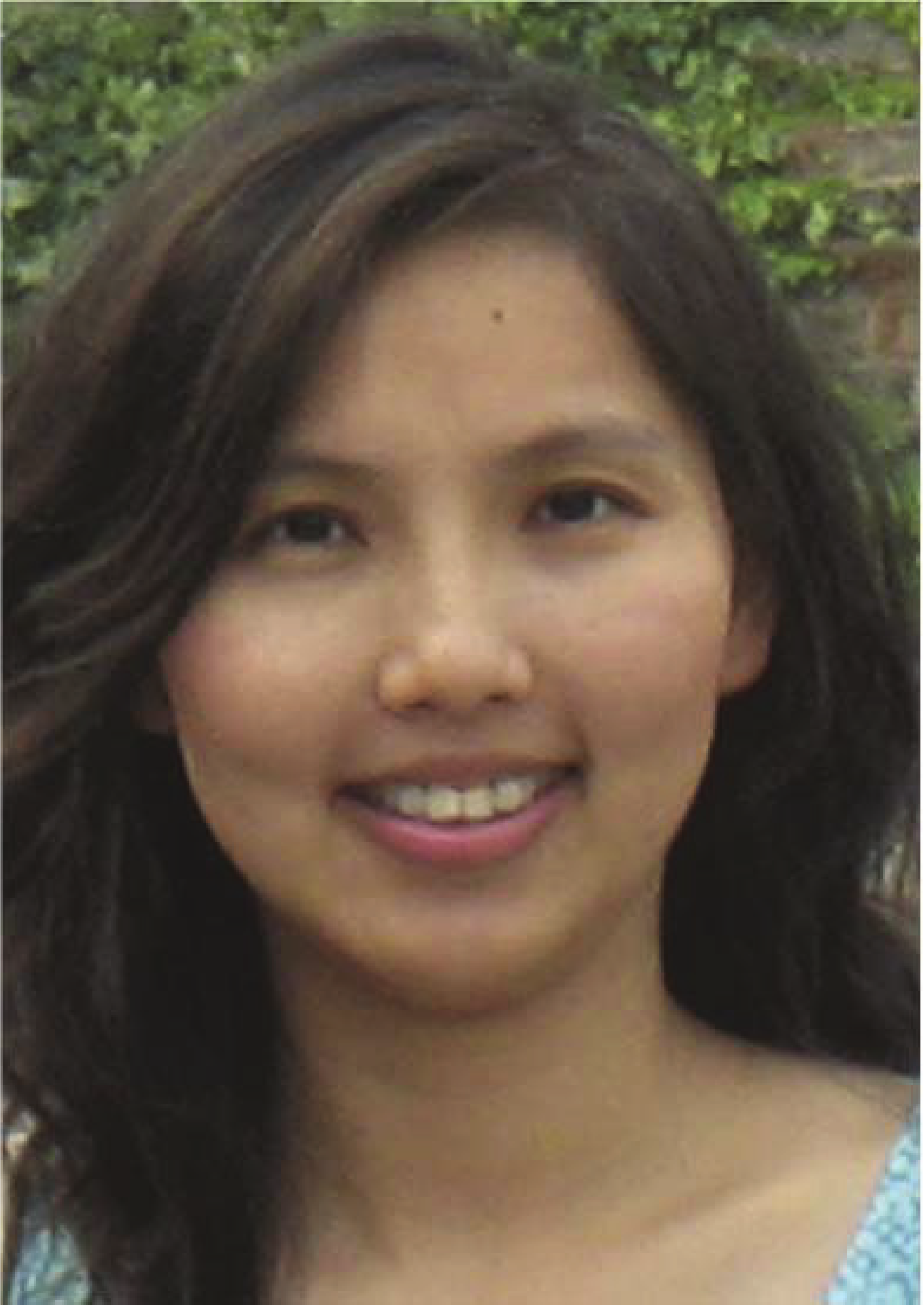}}]{Liuqing Yang}
(S'02-M'04-SM'06-F'15) received the Ph.D. degree from the University of Minnesota, Minneapolis, MN, USA, in 2004. Her main research interests include communications and signal processing. 

Dr. Yang has been actively serving in the technical community, including the organization of many IEEE international conferences, and on the editorial boards of a number of journals, including the IEEE Transactions on Communications, the IEEE Transactions on Wireless Communications, the IEEE Transactions on Intelligent Transportation Systems, and the IEEE Transactions on Signal Processing. She received the Office of Naval Research Young Investigator Program Award in 2007, the National Science Foundation Career Award in 2009, the IEEE GLOBECOM Outstanding Service Award in 2010, the George T. Abell Outstanding Mid-Career Faculty Award and the Art Corey Outstanding International Contributions Award at CSU in 2012 and 2016 respectively, and Best Paper Awards at IEEE ICUWB'06, ICCC'13, ITSC'14, GLOBECOM'14, ICC'16, and WCSP'16.
\end{IEEEbiography}

\end{document}